\documentclass[letterpaper]{article}

\usepackage[T1]{fontenc}

\usepackage[utf8]{inputenc}

\usepackage{placeins}

\usepackage{geometry}
\usepackage{setspace}
\usepackage{amssymb}
\usepackage[utf8]{inputenc}
\usepackage[T1]{fontenc}
\usepackage[style = chem-acs]{biblatex}
\usepackage{graphicx}
\usepackage{float}
\newfloat{scheme}{htbp}{los}
\floatname{scheme}{Scheme}
\floatname{chart}{Chart}
\newfloat{graph}{htbp}{loh}

\usepackage{chemformula} 
\usepackage[version = 4]{mhchem} 

\usepackage[
  colorlinks=false,
  citebordercolor={0 0.65 0},
  linkbordercolor={0 0.65 0},
  urlbordercolor={0 0.65 0},
  pdfborder={0 0 1}
]{hyperref}

\usepackage{authblk}
\author[1]{Tiberius O. Cheche}
\author[1,2]{Yia-Chung Chang*}
\affil[1]{Department of Physics, National Cheng-Kung University, Tainan, Taiwan 704}
\affil[2]{Research Center for Applied Sciences, Academia Sinica, Taipei, Taiwan 11529}

\title{Dielectric function in WSe\textsubscript{2} monolayer Wigner crystal}
\date{*Email: yiachang@gate.sinica.edu.tw}

\begin{document}
\setcounter{secnumdepth}{2}

\maketitle

\begin{abstract}
We develop a Hartree-Fock numerical method for computing the band structure of a two-dimensional Wigner crystal in an electron gas at zero temperature. The ground state is assumed to be fully spin-polarized. Single-particle excitation spectra are evaluated in spin-conserving channel. As an application, we use the developed code to compute the static dielectric function $\epsilon$(\textit{q},0) of a Wigner-crystal state formed in a two-dimensional transition-metal dichalcogenide, specifically monolayer WSe\textsubscript{2}. The dielectric response is obtained from the Hartree-Fock band structure and eigenfunctions through a static Lindhard-type polarizability. The method provides a theoretical tool for investigating screening, band-structure reconstruction, and interaction effects in low-density two-dimensional systems, with possible relevance for future experimental studies.
\end{abstract}

\section{Introduction}
\label{sec:Introd}

The possibility of electron crystallization was first proposed by Wigner in his seminal studies of the interacting electron gas \cite{Wigner1934,Wigner1938}. In the low-density regime, the Coulomb interaction energy dominates over the kinetic energy, and the electron gas can lower its total energy by forming a spatially ordered crystalline state, now known as the Wigner-crystal. Subsequent theoretical work established the basic static and dynamical properties of such ordered electron states. In particular, Bonsall and Maradudin analyzed the classical two-dimensional (2D) Wigner-crystal, showing that the triangular lattice has the lowest energy among the 2D Bravais lattices, and computed its phonon spectrum and long-wavelength dielectric response \cite{Bonsall1977}.

More quantitative quantum descriptions of Wigner crystallization have been developed within both many-body and mean-field frameworks. Hartree--Fock approaches have been used to construct broken-symmetry crystalline solutions, to analyze spin polarization and excitation channels, and to compute the electronic structure of Wigner-crystal states \cite{Trail2003,Bernu2011,JainHuang2025}. Quantum Monte Carlo calculations have provided accurate benchmarks for the ground-state energy and for the liquid--crystal phase boundary in two- and three-dimensional electron gases \cite{Tanatar1989,Drummond2004}.

Recent experimental advances in atomically thin semiconductors have provided direct motivation for applying Wigner-crystal theory to 2D transition-metal dichalcogenide (TMD) materials. Zhou \textit{et al.} reported bilayer Wigner crystals in MoSe$_2$ heterostructures separated by hexagonal boron nitride, without requiring an external magnetic field or a moir\'e potential, and investigated their quantum and thermal melting behavior \cite{Zhou2021}. Smole\'nski \textit{et al.} observed signatures of an electronic Wigner crystal in a pristine monolayer semiconductor at low carrier density, emphasizing the role of large effective mass and reduced dielectric screening in stabilizing zero-field crystallization \cite{Smolenski2021}. More recently, Zhang \textit{et al.} used exciton spectroscopy in monolayer WSe$_2$ to probe Wigner polarons and to distinguish static Umklapp signatures from dynamic polaron resonances \cite{Zhang2025}. In a related moir\'e setting, Tang \textit{et al.} studied bandwidth-controlled metal--insulator transitions in a MoSe$_2$/WS$_2$ superlattice and highlighted dielectric response as a probe of interaction-driven Wigner--Mott physics \cite{Tang2022}.

A 2D Wigner crystal is characterized by a spatially modulated charge density with the periodicity of a triangular lattice. In the present work, we adopt a self-consistent Hartree--Fock approach to compute the single-particle band structure of such a crystalline state and its static dielectric response. The calculation is performed within a restricted Hartree--Fock framework, in which the single-particle orbitals and the charge density are not constrained to preserve the continuous translational symmetry of the uniform electron gas. The self-consistent solution is therefore allowed to develop crystalline density harmonics. In the present implementation, the ground state is assumed to be fully spin-polarized. Consequently, exchange acts only between states with the same spin projection, whereas the opposite-spin particle channel is described by the Hartree contribution evaluated in the frozen Hartree potential generated by the converged spin-polarized Hartree--Fock Wigner-crystal density.

The manuscript is organized as follows. Theory section introduces the Hartree--Fock formulation of the spin-polarized Wigner crystal, its expansion in the plane-wave basis $\left\lvert \mathbf{k}+\mathbf{G} \right\rangle$, and the Lindhard-type expression used to compute the static dielectric response of the insulating crystalline state. Numerical code and results section describes the numerical implementation and presents the calculated band structures and dielectric functions. Conclusions section summarizes the main conclusions. Additional technical details are provided in Supporting Information.



\section{Theory}
\label{sec:theory}

\subsection{Hartree--Fock equation in the $\mathbf{k}+\mathbf{G}$ basis}
\label{sec:hf_kG}

The self-consistent Wigner-crystal state is periodic with respect to the
emergent triangular lattice. The Hartree--Fock (HF) orbitals can therefore be
expanded in Bloch form as
\begin{equation}
\psi_{n\mathbf{k}\sigma}(\mathbf r)=\frac{1}{\sqrt{A}}
\sum_{\mathbf G}
Z_{n\mathbf G\sigma}(\mathbf k)
e^{i(\mathbf k+\mathbf G)\cdot\mathbf r},
\label{eq:bloch_expansion}
\end{equation}
where \(A\) is the normalization area, \(\sigma\) is the spin projection,
\(\mathbf k\) belongs to the first Brillouin zone of the Wigner-crystal
lattice, and \(\mathbf G\) are reciprocal-lattice vectors. Projection onto the
plane-wave basis \(A^{-1/2}e^{i(\mathbf k+\mathbf G)\cdot\mathbf r}\) gives
the matrix eigenvalue problem
\begin{equation}
\sum_{\mathbf G'}
h_{\mathbf G\mathbf G'}^{\sigma}(\mathbf k)
Z_{n\mathbf G'\sigma}(\mathbf k)
=
\varepsilon_{n\mathbf{k}\sigma}
Z_{n\mathbf G\sigma}(\mathbf k),
\label{eq:hf_matrix_eigenproblem}
\end{equation}
with
\begin{equation}
h_{\mathbf G\mathbf G'}^{\sigma}(\mathbf k)
=
T_{\mathbf G\mathbf G'}(\mathbf k)
+
V^H_{\mathbf G\mathbf G'}
-
\Sigma^{F,\sigma}_{\mathbf G\mathbf G'}(\mathbf k).
\label{eq:hf_matrix_decomposition}
\end{equation}

For a parabolic effective-mass band and in the absence of an external
potential, the kinetic contribution is diagonal,
\begin{equation}
T_{\mathbf G\mathbf G'}(\mathbf k)
=
\frac{\hbar^2|\mathbf k+\mathbf G|^2}{2m^*}
\delta_{\mathbf G\mathbf G'},
\label{eq:kinetic_matrix}
\end{equation}
where \(m^*\) is the electron effective mass in the 2D material. The Hartree
term is local and independent of \(\mathbf k\). In reciprocal space,
\begin{equation}
V^H_{\mathbf G\mathbf G'}
=
V_H(\mathbf G-\mathbf G')
=
\widetilde{v}(\mathbf G-\mathbf G')\,
\widetilde{n}(\mathbf G-\mathbf G'),
\label{eq:hartree_matrix}
\end{equation}
where \(\widetilde{v}(\mathbf Q)\) is the Fourier transform of the interaction
potential and \(\widetilde{n}(\mathbf Q)\) is the self-consistent density
Fourier coefficient.

The Fock term is nonlocal and spin diagonal. In the
\(\mathbf{k}+\mathbf{G}\) representation it is
\begin{equation}
\Sigma^{F,\sigma}_{\mathbf G\mathbf G'}(\mathbf k)
=
\sum_{\nu'\in\mathrm{occ}}
\sum_{\mathbf k'}
w_{\mathbf k'}
\sum_{\mathbf G_1}
Z_{\nu'\mathbf G_1\sigma}(\mathbf k')
Z^{*}_{\nu',\mathbf G_1-\mathbf G+\mathbf G',\sigma}(\mathbf k')
\,
\widetilde{v}
\!\left[
(\mathbf k+\mathbf G)-(\mathbf k'+\mathbf G_1)
\right],
\label{eq:fock_matrix}
\end{equation}
where \(w_{\mathbf k'}\) are the Brillouin-zone integration weights and
\(\nu'\) labels the occupied HF bands entering the spin-resolved density
matrix. Terms for which the shifted vector
\(\mathbf G_1-\mathbf G+\mathbf G'\) lies outside the retained plane-wave
basis are omitted. In the fully spin-polarized one-band calculation considered
here, only one spin sector is occupied and the occupied-band sum reduces to
\(\nu'=1\); consequently, exchange contributes only to the same-spin particle
channel. The coordinate-space HF equation and the derivation of the matrix
elements, including the reciprocal-lattice selection rule implicit in
Eq.~\eqref{eq:fock_matrix}, are given in \ref{sec:si_hf_matrix_elements}.

For the band-structure calculations reported below, the target momenta $\mathbf k$ are taken in the irreducible $\Gamma$--$K$--$M$ wedge, while the source sum over $\mathbf k'$ in Eq.~\eqref{eq:fock_matrix} is evaluated explicitly over the full Brillouin-zone mesh reconstructed from the wedge solution. Thus the exchange field is built from the full Brillouin zone rather than from a symmetry-reduced source sum. The coefficients $Z_{n\mathbf G\sigma}(\mathbf k)$ are obtained self-consistently: at a given iteration, the density harmonics $\eta_W(\mathbf Q)$ define the Hartree contribution, whereas the occupied eigenvectors from the previous iteration define the Fock contribution.

The effective electrostatic interaction between charge carriers confined to a 2D semiconductor can be described by the Keldysh interaction \cite{Keldysh1979,Mostaani2017,Danovich2018}. In reciprocal space, for an isotropic dielectric environment, we write
\begin{equation}
\widetilde{v}_{\mathrm{K}}(q)=
\frac{2\pi e^2}{\varepsilon q\left(1+q\rho_0\right)} .
\label{eq:keldysh_interaction}
\end{equation}
Here, $q$ denotes the wave-vector length, $\varepsilon$ is the effective dielectric constant of the medium surrounding the monolayer, and $\rho_0$ is the Keldysh screening length. In the convention used here, $\rho_0=\kappa/(2\varepsilon)$, where $\kappa$ denotes the in-plane polarizability of the 2D material.

After diagonalizing Eq.~\eqref{eq:hf_matrix_eigenproblem} for all $\mathbf k$ points, the new eigenvectors are used to reconstruct the density harmonics according to
\begin{equation}
\eta_W(\mathbf Q)
=
\frac{1}{A_W}
\sum_{\nu\in\mathrm{occ}}
\sum_{\mathbf k}
w_{\mathbf k}
\sum_{\mathbf G}
Z^*_{\nu\mathbf G\sigma}(\mathbf k)
Z_{\nu,\mathbf G+\mathbf Q,\sigma}(\mathbf k),
\label{eq:etaW_from_Z}
\end{equation}
where $A_W$ is the Wigner-crystal primitive-cell area and the Brillouin-zone weights are normalized as $\sum_{\mathbf k}w_{\mathbf k}=1$. The index $\nu$ is a dummy occupied-band index used in the density reconstruction. In the present one-band spin-polarized implementation, the sum over occupied bands contains only the lowest occupied band and as mentioned $\nu=1$. The numerical implementation uses an irreducible wedge of the Brillouin zone, but the density reconstruction is equivalent to a full-zone summation after applying the symmetry images of the eigenvectors. The self-consistency loop is therefore closed by repeatedly updating the pair $\{Z_{n\mathbf G\sigma}(\mathbf k),\eta_W(\mathbf Q)\}$ until convergence.

The self-consistent Hartree--Fock cycle is initialized from a periodic charge density with the periodicity of the Wigner-crystal lattice. Since the density is periodic over the Wigner-Seitz cell of area $A_W$, it can be expanded in reciprocal-lattice harmonics,
\begin{equation}
n^{(0)}(\mathbf r)=\sum_{\mathbf Q}
\eta_W^{(0)}(\mathbf Q)
e^{i\mathbf Q\cdot\mathbf r},
\label{eq:density_fourier_eta}
\end{equation}
where \(\mathbf Q=m\mathbf b_1+n\mathbf b_2\), with \(m,n\in\mathbb Z\), denotes a reciprocal-lattice vector. For one electron per primitive cell, the normalization condition is
\(\eta_W^{(0)}(\mathbf 0)=1/A_W\).

For the initial state, we use localized Gaussian-like Bloch amplitudes in the plane-wave basis $\left\lvert \mathbf k+\mathbf G\right\rangle$. For each wave vector $\mathbf k$ in the Brillouin-zone and $\mathbf G$ reciprocal lattice vector, the initial coefficients are taken as
\begin{equation}
F_{\mathbf k}^{(0)}(\mathbf G)=\frac{
\exp\left[-\left|\mathbf k+\mathbf G\right|^2/(2\beta)\right]
}{
\left[
\sum_{\mathbf G'}
\exp\left[-\left|\mathbf k+\mathbf G'\right|^2/\beta\right]
\right]^{1/2}
},
\label{eq:gaussian_initial_coeff}
\end{equation}
where $\beta$ controls the width of the initial packet in reciprocal space. 
In the numerical implementation, the input parameter is the dimensionless Gaussian width
$\beta_0=\beta (a_B^*)^2$, or equivalently $\beta=\beta_0/(a_B^*)^2$, where $\beta$ is measured in nm$^{-2}$. The effective Bohr radius is defined as $a_B^*=(\varepsilon m_0/m^*)a_0$, with $a_0$ the Bohr radius in vacuum, $m_0$ the free-electron mass, and $\varepsilon$ the effective dielectric constant of the medium surrounding the monolayer.


The corresponding initial density harmonics are computed from the plane-wave coefficients as
\begin{equation}
\eta_W^{(0)}(\mathbf Q)
=
\frac{1}{A_W}
\sum_{\mathbf k}
w_{\mathbf k}
\sum_{\mathbf G}
F_{\mathbf k}^{(0)}(\mathbf G)
F_{\mathbf k}^{(0)}(\mathbf G+\mathbf Q),
\label{eq:initial_etaW}
\end{equation}
where the weights satisfy $\sum_{\mathbf k}w_{\mathbf k}=1$. For a uniform full-BZ mesh, $w_{\mathbf k}=1/N_k$, and Eq.~\eqref{eq:initial_etaW} reduces to the explicit form with the prefactor $1/(A_WN_k)$. This construction guarantees the correct normalization condition,  while providing a periodic crystalline initial guess for the subsequent self-consistent Hartree--Fock iterations. 
Proof of eqs. (8) and (11) is provided in \ref{Ini_dens}.

\subsection{Static dielectric function}
\label{sec:static_dielectric_function}


The dielectric response of a Wigner-crystal can be discussed from complementary viewpoints. In the classical long-wavelength limit, the response is governed by the collective elastic and phonon modes of the electron lattice, as in the analysis of Bonsall and Maradudin for the 2D Wigner-crystal \cite{Bonsall1977}. Screening and capacitive response have also been used experimentally as probes of Wigner-crystal melting and dynamics in high-mobility two-dimensional electron systems \cite{Deng2019,Zhao2023}. The present calculation addresses a different, microscopic contribution: the static density response obtained from the self-consistent Hartree--Fock quasiparticle bands and eigenvectors of the crystalline state.

The formula used below is the insulating, folded-Brillouin-zone analogue of the independent-particle polarizability underlying the Lindhard response of the electron gas \cite{Lindhard1954} and the Stern polarizability of the homogeneous two-dimensional electron gas \cite{Stern1967}. Because the Wigner crystal is a periodic broken-symmetry state, the relevant matrix elements are those of a band problem in the Wigner-crystal Brillouin zone, closely analogous to the Adler--Wiser formulation of the dielectric response of periodic solids \cite{Adler1962,Wiser1963}. In this work we evaluate the scalar longitudinal static response using the Hartree--Fock eigenvalues and plane-wave coefficients in the $\mathbf k+\mathbf G$ basis.

The converged Hartree--Fock eigenvalues and eigenvectors are used to compute the static density response of the insulating Wigner-crystal state. We consider a scalar longitudinal perturbation with wave vector $\mathbf q$. For each initial momentum $\mathbf k$ in the first Wigner-crystal Brillouin zone, the final momentum is folded back to the same zone according to
\begin{equation}
\mathbf k_{\mathbf q}
=
\mathbf k+\mathbf q-\mathbf G_{\mathrm f}(\mathbf k,\mathbf q),
\qquad
\mathbf k_{\mathbf q}\in \mathrm{1BZ},
\label{eq:kq_fold_main}
\end{equation}
where $\mathbf G_{\mathrm f}(\mathbf k,\mathbf q)$ is a reciprocal-lattice vector of the Wigner-crystal. The corresponding density matrix element is
\begin{equation}
M_{nm}(\mathbf k,\mathbf q)
=
\left\langle
n\mathbf k
\left|
e^{-i\mathbf q\cdot\mathbf r}
\right|
m\mathbf k_{\mathbf q}
\right\rangle
=
\sum_{\mathbf G}
Z^*_{n\mathbf G}(\mathbf k)
Z_{m,\mathbf G+\mathbf G_{\mathrm f}}(\mathbf k_{\mathbf q}) .
\label{eq:density_form_factor_main}
\end{equation}
Terms for which $\mathbf G+\mathbf G_{\mathrm f}$ lies outside the retained $G$ basis are omitted.

At zero temperature and for a fully spin-polarized insulating state, the static Hartree--Fock band polarizability is evaluated as
\begin{equation}
\Pi_{\mathrm{HF}}(\mathbf q,0)
=
\frac{1}{A_W}
\sum_{\mathbf k}
w_{\mathbf k}
\sum_{n\in\mathrm{occ}}
\sum_{m\in\mathrm{emp}}
\left|
M_{nm}(\mathbf k,\mathbf q)
\right|^2
\frac{
\Delta E_{mn}(\mathbf k,\mathbf q)
}{
\Delta E_{mn}^2(\mathbf k,\mathbf q)+\eta_{\Pi}^2
},
\label{eq:PiHF_static_main}
\end{equation}
where
\begin{equation}
\Delta E_{mn}(\mathbf k,\mathbf q)
=
\varepsilon_{m\mathbf k_{\mathbf q}}
-
\varepsilon_{n\mathbf k},
\label{eq:DeltaE_main}
\end{equation}
and $\eta_{\Pi}$ is a small numerical broadening. In the formal static limit one takes $\eta_{\Pi}\rightarrow0^+$, so that Eq.~\eqref{eq:PiHF_static_main} reduces to the usual interband Lindhard denominator $1/\Delta E_{mn}$.

The static dielectric function is then defined as
\begin{equation}
\varepsilon_{\mathrm{HF}}(\mathbf q,0)
=
1+
\widetilde v_{\mathrm K}(q)
\Pi_{\mathrm{HF}}(\mathbf q,0),
\qquad
q=|\mathbf q| ,
\label{eq:epsilonHF_static_main}
\end{equation}
with $\widetilde v_{\mathrm K}(q)$ given by Eq.~\eqref{eq:keldysh_interaction}. No additional spin-degeneracy factor is included, because the calculation is performed for a single fully spin-polarized occupied sector.

The insulating character of the self-consistent Wigner crystal fixes the long-wavelength limit. At $\mathbf q=\mathbf0$, orthonormality gives $M_{nm}(\mathbf k,\mathbf0)=\delta_{nm}$, so the occupied--empty interband contribution vanishes and $\Pi_{\mathrm{HF}}(\mathbf0,0)=0$. For small nonzero $\mathbf q$, the interband form factor is linear in $q$, hence $\Pi_{\mathrm{HF}}(\mathbf q,0)\propto q^2$. Since the two-dimensional Keldysh interaction behaves as $\widetilde v_{\mathrm K}(q)\propto1/q$ for $q\rightarrow0$, Eq.~\eqref{eq:epsilonHF_static_main} gives
\begin{equation}
\varepsilon_{\mathrm{HF}}(\mathbf0,0)=1,
\qquad
\lim_{q\rightarrow0}\varepsilon_{\mathrm{HF}}(\mathbf q,0)=1 .
\label{eq:epsilon_q0_continuity_main}
\end{equation}
The derivation of the form factor and of the small-$q$ limit is given in \ref{sec:si_static_polarizability}

\subsection{Triangular Wigner-crystal lattice and Brillouin-zone sampling}
\label{sec:tmd_wc_lattice}

For an isotropic two-dimensional material with repulsive Coulomb-like interactions, the energetically favored Bravais lattice is triangular \cite{Bonsall1977}. This motivates the use of a triangular Wigner-crystal primitive cell for the present TMD monolayer calculation, where low carrier density, large effective mass, and reduced dielectric screening provide a favorable setting for Wigner crystallization \cite{Smolenski2021}.

We denote the real-space primitive vectors of the Wigner crystal by $\mathbf a_1$ and $\mathbf a_2$ and choose their orientation such that the Wigner-crystal primitive-cell area is
$A_W = \hat{\mathbf z} \cdot
\left(
\mathbf a_1\times\mathbf a_2
\right)>0 .$
The corresponding reciprocal primitive vectors are
\(\mathbf b_1=(2\pi/A_W)(\mathbf a_2\times\hat{\mathbf z})\) and
\(\mathbf b_2=(2\pi/A_W)(\hat{\mathbf z}\times\mathbf a_1)\).
A general reciprocal-lattice vector is then written in Miller form as
$\mathbf G_{mn}=m\mathbf b_1+n\mathbf b_2 ,
\qquad
m,n\in\mathbb Z$.

We use this hexagonal Brillouin zone as the fundamental momentum-space cell for the Hartree--Fock band structure. The standard high-symmetry points are denoted by $\Gamma$ at the zone center, $M$ at the midpoint of a hexagon edge, and $K$ at a hexagon corner. The single-particle states are labeled by a reduced crystal momentum $\mathbf k$ inside this first Brillouin zone and by reciprocal-lattice vectors $\mathbf G$, leading to the $\mathbf k+\mathbf G$ plane-wave representation used below. The Brillouin-zone sampling is performed in the irreducible $\Gamma$--$K$--$M$ wedge of the hexagonal zone as shown in Figure 1.

\begin{figure}[t]
\centering
\includegraphics[width=0.4\linewidth]{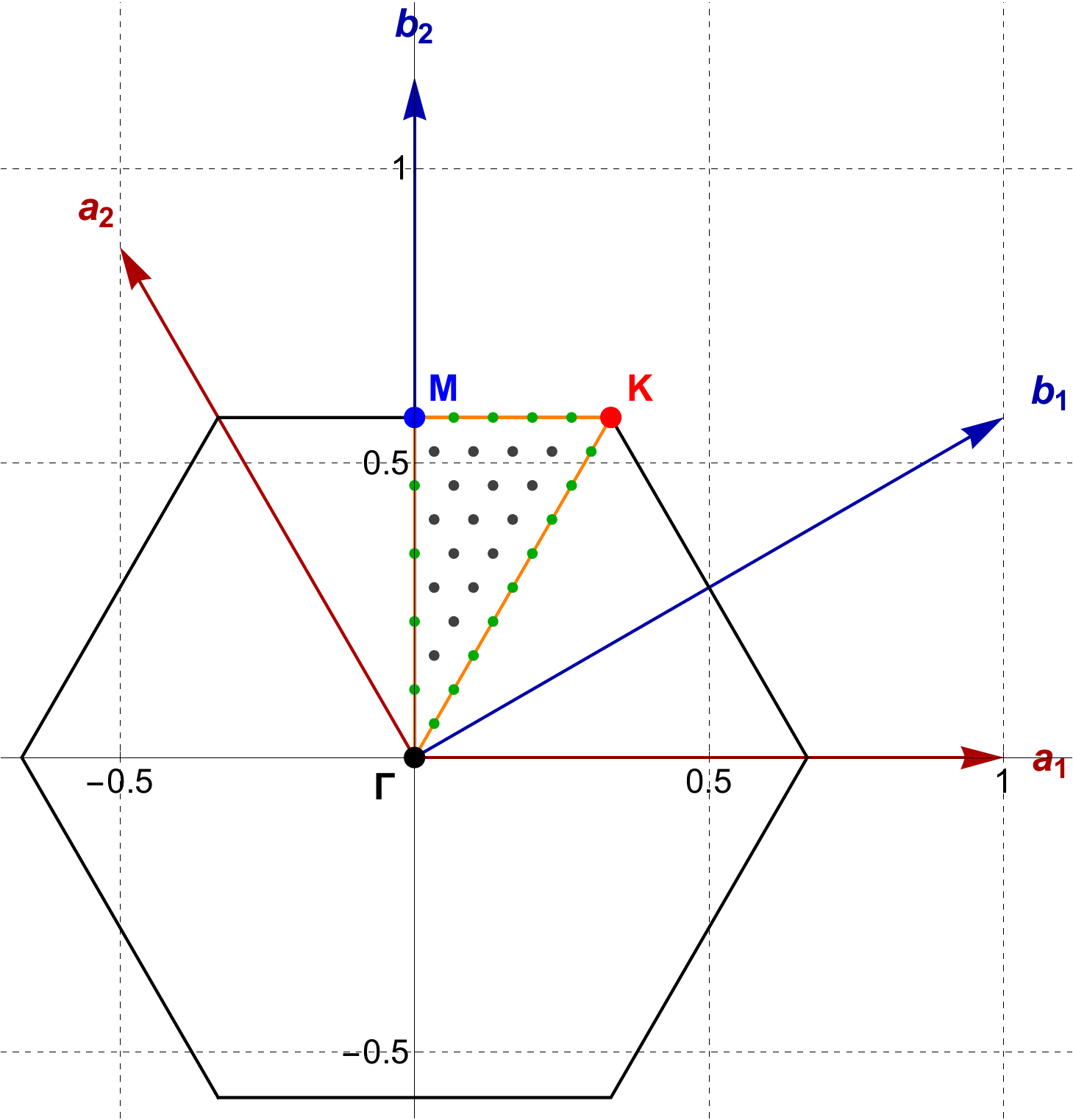}
\caption{Hexagonal first Brillouin zone of the triangular Wigner-crystal lattice. The irreducible $\Gamma$--$K$--$M$ wedge used for the Brillouin-zone sampling is indicated.}
\label{fig:bz_wedge}
\end{figure}

The plane-wave basis is truncated by retaining a symmetry-compatible star of reciprocal vectors. The retained set is selected by a using a circular cutoff in Cartesian reciprocal space,
$(\mathcal G_{\mathrm{cut}}={\mathbf G_{mn}:|\mathbf G_{mn}|^2\leq G_{\mathrm{cut}}^2}).$
This radial cutoff makes the accepted set a union of complete stars under the point group $D_6$ of the hexagonal Brillouin zone. Further details on the geometry of the triangular Wigner-crystal lattice are provided in
Sec.~\ref{sec:si_miller_map}.


The Hartree--Fock band structure is obtained by diagonalizing the matrix Hamiltonian at the representative $\mathbf k$ points of the irreducible $\Gamma$--$K$--$M$ wedge. For the exchange term, however, the source momentum sum is evaluated over the full Brillouin-zone mesh. This full mesh is generated from the wedge representatives by applying the distinct symmetry images. 
Thus the target momenta are represented in the irreducible wedge, while the exchange field is evaluated by an explicit full-zone summation. A symmetry-reduced wedge form of the same exchange sum is possible and is described in ~\ref{sec:Full_zone}, but it is not used in the production results reported here.

\section{Numerical code and results}
\label{sec:numerical_results}

\subsection{Numerical workflow and convergence}
\label{sec:numerical_workflow}

The reciprocal-lattice basis is selected by a circular cutoff in $|\mathbf G|^2$, which preserves complete $D_6$ stars of the triangular Wigner-crystal reciprocal lattice. The Brillouin-zone sampling is performed in the irreducible $\Gamma$--$K$--$M$ wedge, while full-zone symmetry images are reconstructed when required for the exchange matrix and for the dielectric form factors. The code supports Coulomb, Keldysh, and Stern-family interaction kernels; the production results reported below use the Keldysh interaction appropriate for 2D TMD monolayer.
A special numerical treatment is required for the singular $q=0$ component of the Coulomb and Keldysh interactions entering the Hartree--Fock matrix elements. In the production calculations we replace the singular value by the analytic average of the interaction over a small circular cell of radius $q_c$ around the origin in reciprocal space,
\begin{equation}
\overline{v}(0)
=
\frac{1}{\pi q_c^2}
\int_{|\mathbf q|\le q_c}
d^2q\,
\widetilde v(q).
\label{eq:qcell_average_definition}
\end{equation}
For the Keldysh interaction in Eq.~\eqref{eq:keldysh_interaction}, the cell average gives
\begin{equation}
\overline{v}_K(0)
=
\frac{4\pi e^2}{\varepsilon q_c^2\rho_0}
\ln(1+\rho_0 q_c).
\label{eq:keldysh_qcell_average}
\end{equation}
For $q>q_c$, the unmodified interaction $\widetilde v_K(q)$ is used. This replacement is applied only to the singular Coulomb/Keldysh kernels. A genuine Stern-screened interaction is finite at $q=0$ and is therefore evaluated directly.

The self-consistent cycle proceeds as follows. Starting from a periodic Gaussian density with the Wigner-crystal periodicity, the Hartree matrix is built from the density harmonics $\eta_W(\mathbf Q)$. After a short Hartree warm-up, the exchange matrix is evaluated from the occupied Hartree--Fock eigenvectors and the matrix Hamiltonian is diagonalized at each representative wedge momentum. The updated eigenvectors are then used to reconstruct the density harmonics, closing the self-consistency loop. Convergence is monitored through the change of $\eta_W(\mathbf Q)$, the stability of the total energy, and the regularity of the band structure at high-symmetry points.

The main numerical parameters are summarized in Table~\ref{tab:numerical_parameters}
in Sec.~\ref{sec:si_miller_map}.
The convergence of the results was checked against the number of retained reciprocal vectors, the wedge mesh density, the number of empty bands retained in the polarizability, and the broadening parameter $\eta_{\Pi}$ entering Eq.~\eqref{eq:PiHF_static_main}.


The Hartree-only part of the self-consistent problem can possess more than one numerical attractor for the same initial parameter set. This is expected in a broken-symmetry calculation, where different preconditioned density patterns may be stabilized before exchange is activated. In the production calculation, the Hartree warm-up is used only as a stabilizing stage for the subsequent Hartree--Fock iteration, not as an independent physical solution. When more than one Hartree-preconditioned branch was obtained, the branch retained for the Hartree--Fock calculation was selected by three criteria: preservation of the triangular Wigner-crystal symmetry in the density, regularity of the band structure after exchange is included, and recovery of the expected degeneracy pattern at the high-symmetry points of the hexagonal Brillouin zone. The alternative branch was discarded because it led to an irregular Hartree--Fock spectrum and to a density pattern inconsistent with the intended triangular Wigner-crystal solution.

\subsection{Band, density, and dielectric signatures of Wigner-crystal order}
\label{sec:hf_wc_signatures}

We characterize the converged Hartree--Fock solution using three complementary
signatures of Wigner-crystal order: the symmetry structure of the band spectrum,
the amplitude of the first density harmonics, and the static dielectric response.
These quantities are not used as a thermodynamic proof of a liquid--crystal
transition, but as phase-sensitive indicators of the crystalline character of
the self-consistent solution at a given $r_s$.

The calculated band structure provides a sensitive spectral signature of
triangular Wigner-crystal order. At the high-symmetry points $\Gamma$, $K$, and
$M$, the Hartree--Fock bands are constrained by the little groups of the
corresponding momenta and are expected to exhibit the associated degeneracy
structure and regularity. As the density increases, equivalently as $r_s$
decreases, the system approaches the weak-crystal or liquid side. In this
regime, the self-consistent crystalline solution is expected to become less
robust, and the high-symmetry band signatures may become less sharply defined.

The strength of the Wigner-crystal density modulation is estimated by the
first-star order parameter
\begin{equation}
M_1
=
\frac{1}{|\eta_W(\mathbf 0)|}
\frac{1}{6}
\sum_{\mathbf Q\in \mathcal S_1}
\left|
\eta_W(\mathbf Q)
\right| ,
\label{eq:M1_order_parameter}
\end{equation}
where $\eta_W(\mathbf Q)$ are the Fourier coefficients of the charge density
and $\mathcal S_1$ denotes the six shortest nonzero reciprocal-lattice vectors
of the triangular Wigner crystal. The normalization by $|\eta_W(\mathbf 0)|$
removes the overall density scale, so that $M_1$ measures the relative
amplitude of the fundamental density modulation. The first-star harmonics are
the most relevant components for characterizing crystalline order because they
are the longest-wavelength Fourier components compatible with the triangular
broken-symmetry pattern. They therefore appear first at the onset of density
modulation and are less sensitive than higher harmonics to short-range details
of the charge profile. Accordingly, $M_1\simeq0$ signals a nearly uniform or
weakly modulated density, whereas a finite $M_1$ indicates a
Wigner-crystal-like density modulation.

The static dielectric function provides a complementary response signature of
the converged state. It is evaluated from the Hartree--Fock band polarizability
introduced in Sec.~\ref{sec:static_dielectric_function}. For each wave vector
$\mathbf q$, the numerical implementation folds $\mathbf k+\mathbf q$ back to
the first Wigner-crystal Brillouin zone, evaluates the density form factor
$M_{nm}(\mathbf k,\mathbf q)$ from the converged eigenvectors, and sums over
occupied--empty band pairs. The same reconstructed full-zone representation
used in the exchange calculation is used to ensure that the folding operation
and the $G$-index shift in the form factor are treated consistently. The resulting dielectric function is checked against the long-wavelength
constraints derived in Sec.~\ref{sec:static_dielectric_function},
$\varepsilon_{\mathrm{HF}}(\mathbf0,0)=1$ and
$\lim_{q\to0}\varepsilon_{\mathrm{HF}}(\mathbf q,0)=1$, which test the
consistency of the numerical folding procedure and occupied--empty band
summation.

\subsection{Spin channels and exchange scaling}
\label{sec:spin_channels_exchange_scaling}

The self-consistent calculation used for the main Wigner-crystal solution is fully spin polarized. In this case the occupied background contains a single spin component and the same-spin particle spectrum is obtained from
\begin{equation}
h^{\mathrm{SC}}_{\mathbf G\mathbf G'}(\mathbf k)
=
T_{\mathbf G\mathbf G'}(\mathbf k)
+
V^H_{\mathbf G\mathbf G'}
-
X_{\mathbf G\mathbf G'}(\mathbf k),
\label{eq:sc_channel_hamiltonian}
\end{equation}
where $X_{\mathbf G\mathbf G'}$ is the positive exchange matrix built from the occupied spin-polarized Hartree--Fock state. The spin-reversed particle spectrum is obtained in the frozen Hartree potential of the same converged density, but without exchange with the occupied background,
($\Sigma^{F,\sigma}_{\mathbf G\mathbf G'}(\mathbf k)\to 0$ in
Eq.~\eqref{eq:sc_channel_hamiltonian}).

For comparison, a restricted unpolarized estimate can be generated by scaling
the exchange matrix by one half,
\begin{equation}
h^{\mathrm{unpol}}_{\mathbf G\mathbf G'}(\mathbf k)
\approx
T_{\mathbf G\mathbf G'}(\mathbf k)
+
V^H_{\mathbf G\mathbf G'}
-
\frac{1}{2}
X_{\mathbf G\mathbf G'}(\mathbf k).
\label{eq:unpolarized_estimate_hamiltonian}
\end{equation}
This approximation corresponds to assuming equal spin densities,
$\rho_\uparrow=\rho_\downarrow=\rho/2$, while keeping the same total Hartree
density. It should not be confused with a fully self-consistent two-spin
Hartree--Fock calculation.

In the present work we restrict the quantitative analysis to the fully
spin-polarized case, for which the Hartree--Fock exchange field is treated
self-consistently within a single occupied spin sector. The same formulation
also provides a controlled way to analyze spin-reversed frozen-density
excitations, where the Hartree potential is kept fixed and the same-spin exchange term is absent. By contrast, a genuine unpolarized Wigner crystal
requires a self-consistent two-spin Hartree--Fock treatment, including the
coupled evolution of the two spin densities and their same-spin exchange
fields. This more complex case is currently under investigation and will be addressed separately.

\subsection{Band-structure results}
\label{sec:band_structure_results}

\begin{figure*}[t]
\centering
\setlength{\tabcolsep}{1pt}
\renewcommand{\arraystretch}{1.0}

\begin{tabular}{c c c c}
\textbf{Initial density}
&
\textbf{Converged density}
&
\textbf{Band structure}
&
\textbf{$\varepsilon_{\mathrm{HF}}(q,0)$}
\\[0.4ex]

\begin{tabular}{c}
\includegraphics[width=0.238\textwidth]{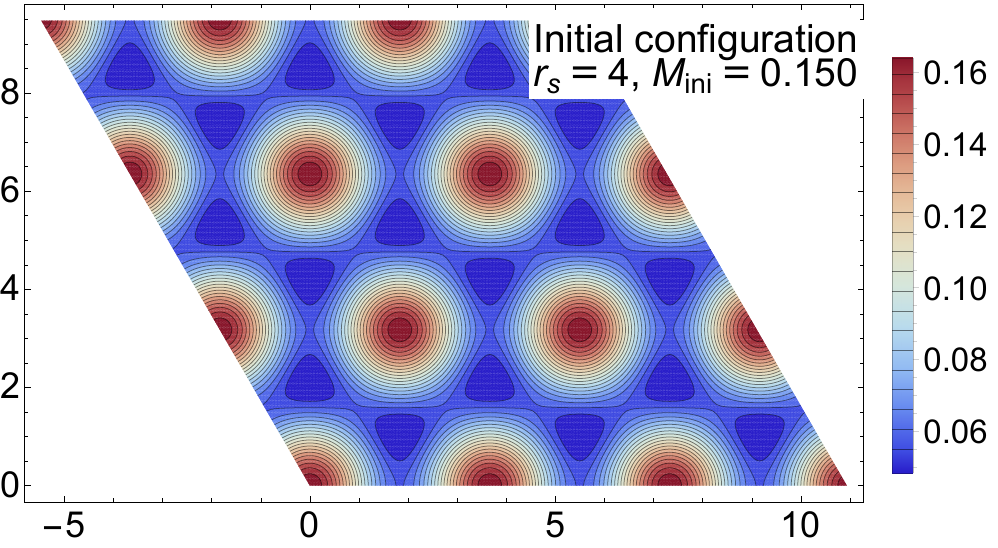}\\[-0.2ex]
\textbf{(a)}
\end{tabular}
&
\begin{tabular}{c}
\includegraphics[width=0.238\textwidth]{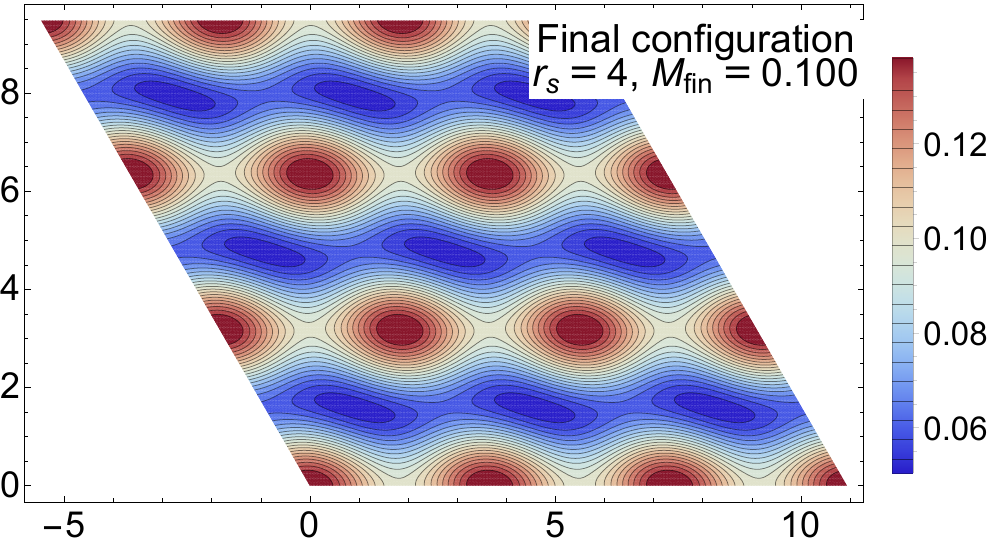}\\[-0.2ex]
\textbf{(b)}
\end{tabular}
&
\begin{tabular}{c}
\includegraphics[width=0.238\textwidth]{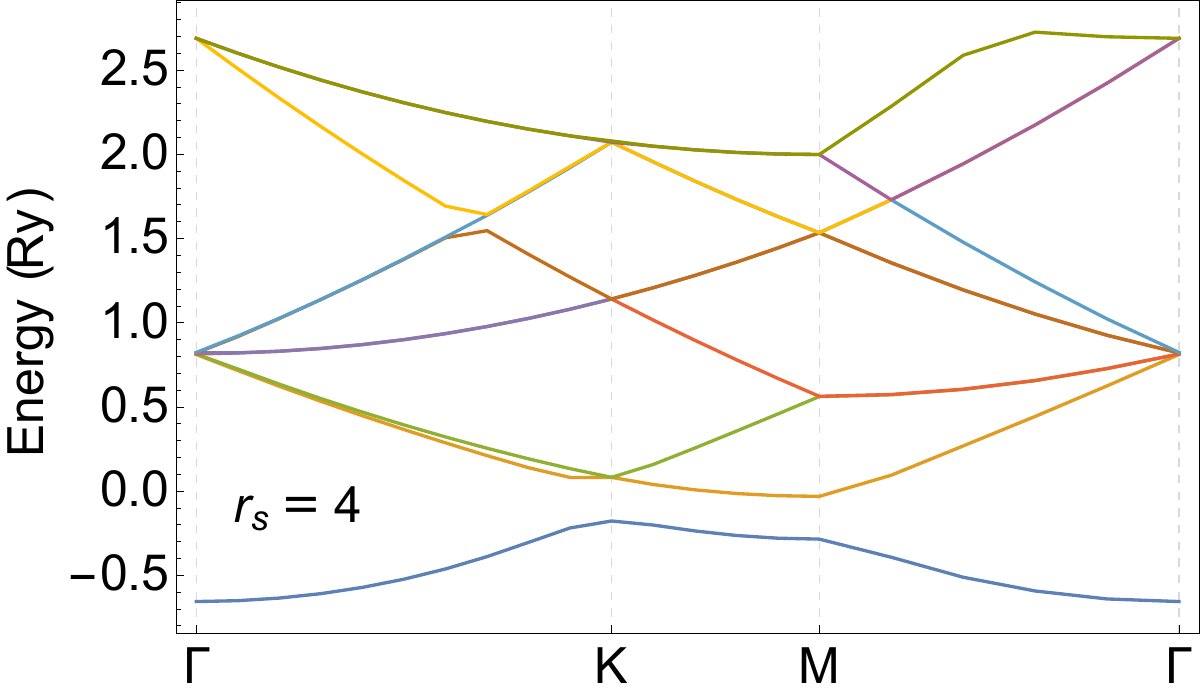}\\[-0.2ex]
\textbf{(c)}
\end{tabular}
&
\begin{tabular}{c}
\includegraphics[width=0.238\textwidth]{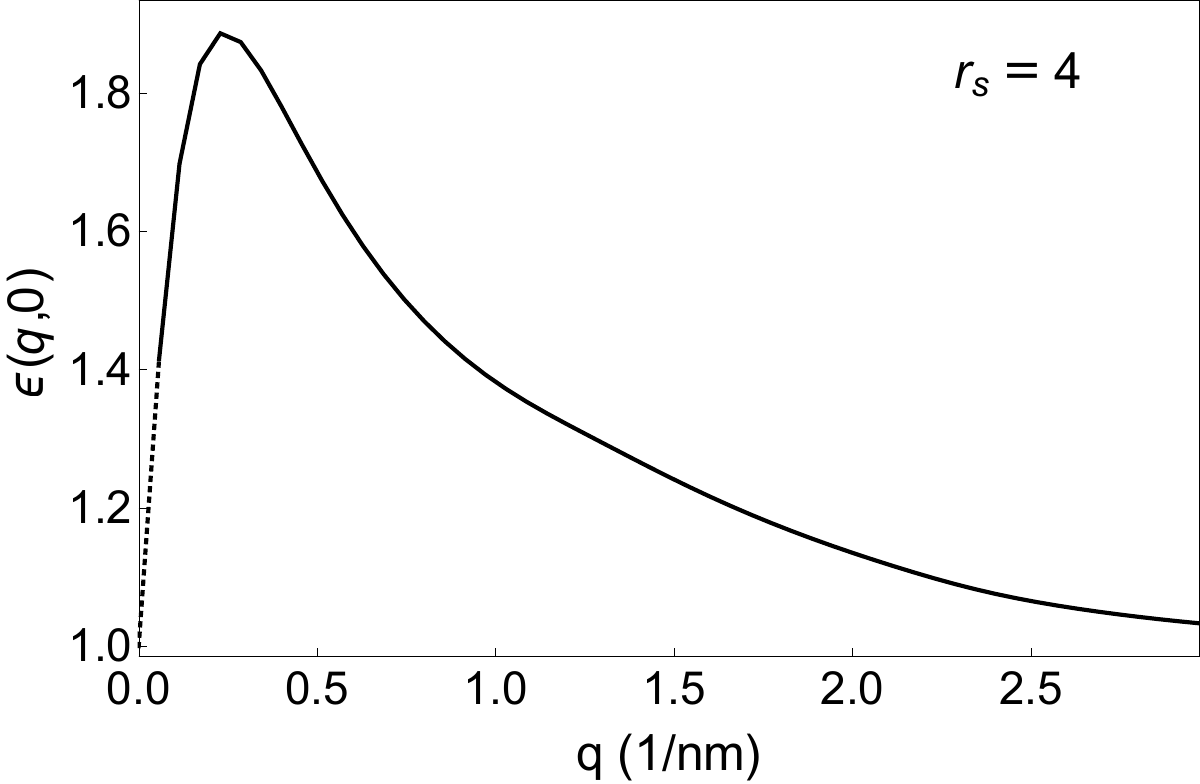}\\[-0.2ex]
\textbf{(d)}
\end{tabular}
\\[0.3ex]

\begin{tabular}{c}
\includegraphics[width=0.238\textwidth]{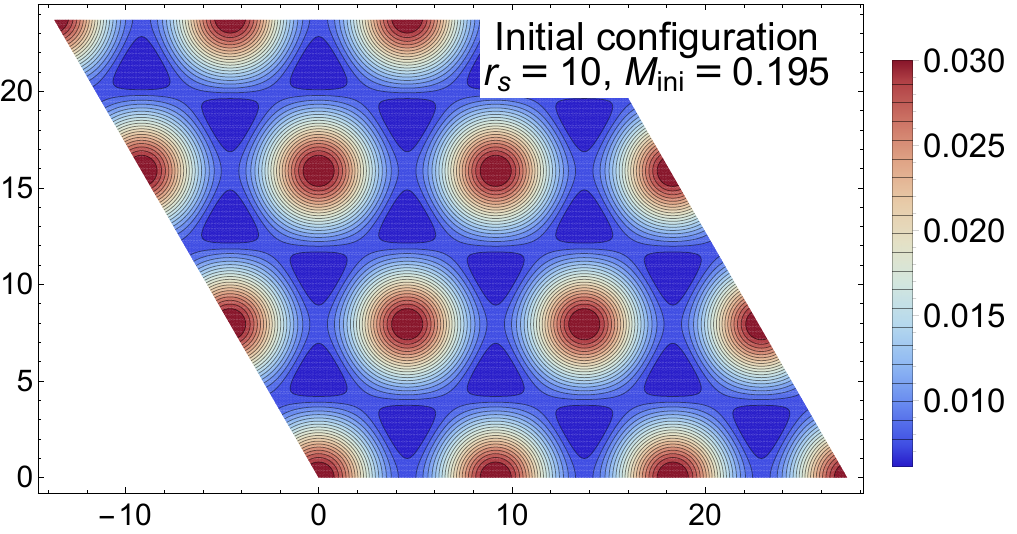}\\[-0.2ex]
\textbf{(e)}
\end{tabular}
&
\begin{tabular}{c}
\includegraphics[width=0.238\textwidth]{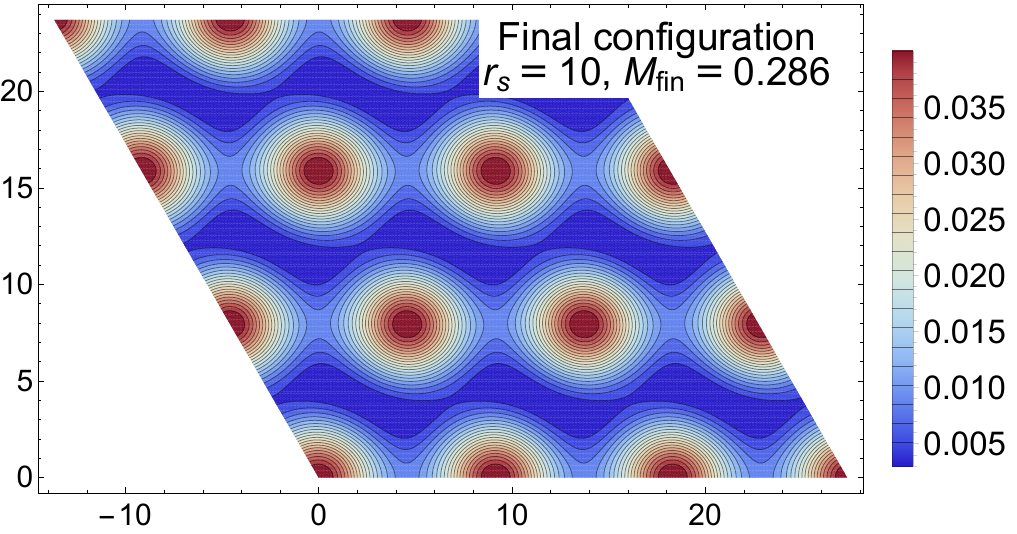}\\[-0.2ex]
\textbf{(f)}
\end{tabular}
&
\begin{tabular}{c}
\includegraphics[width=0.238\textwidth]{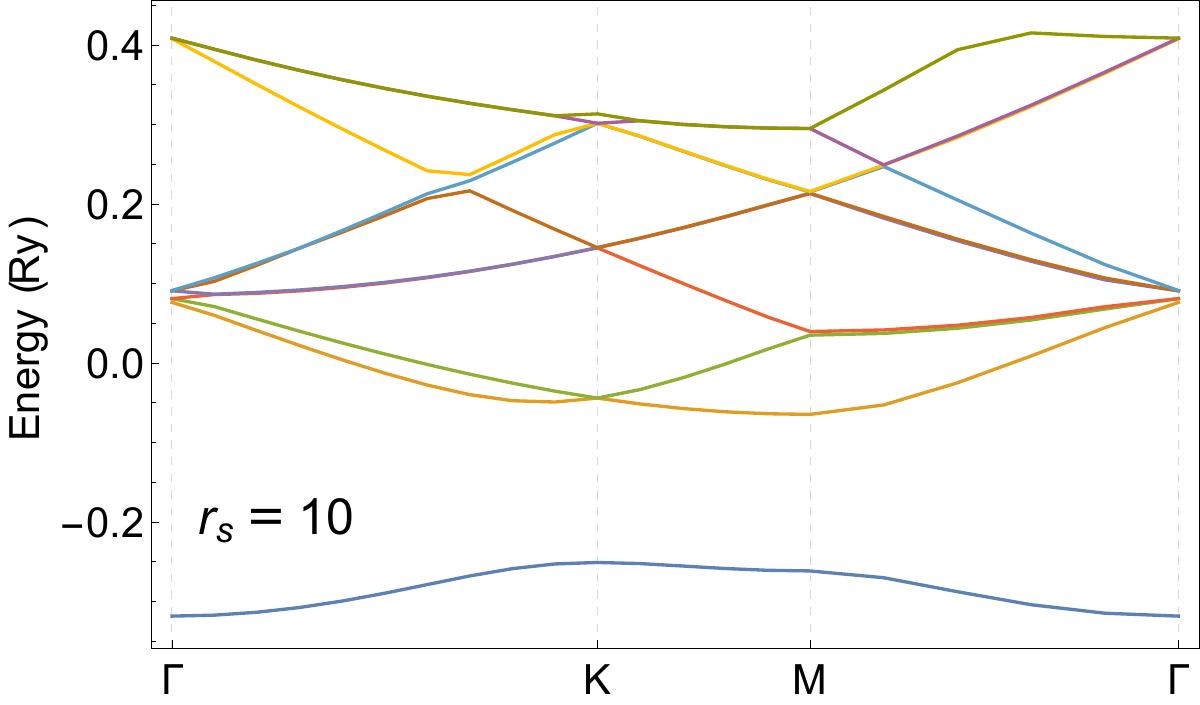}\\[-0.2ex]
\textbf{(g)}
\end{tabular}
&
\begin{tabular}{c}
\includegraphics[width=0.238\textwidth]{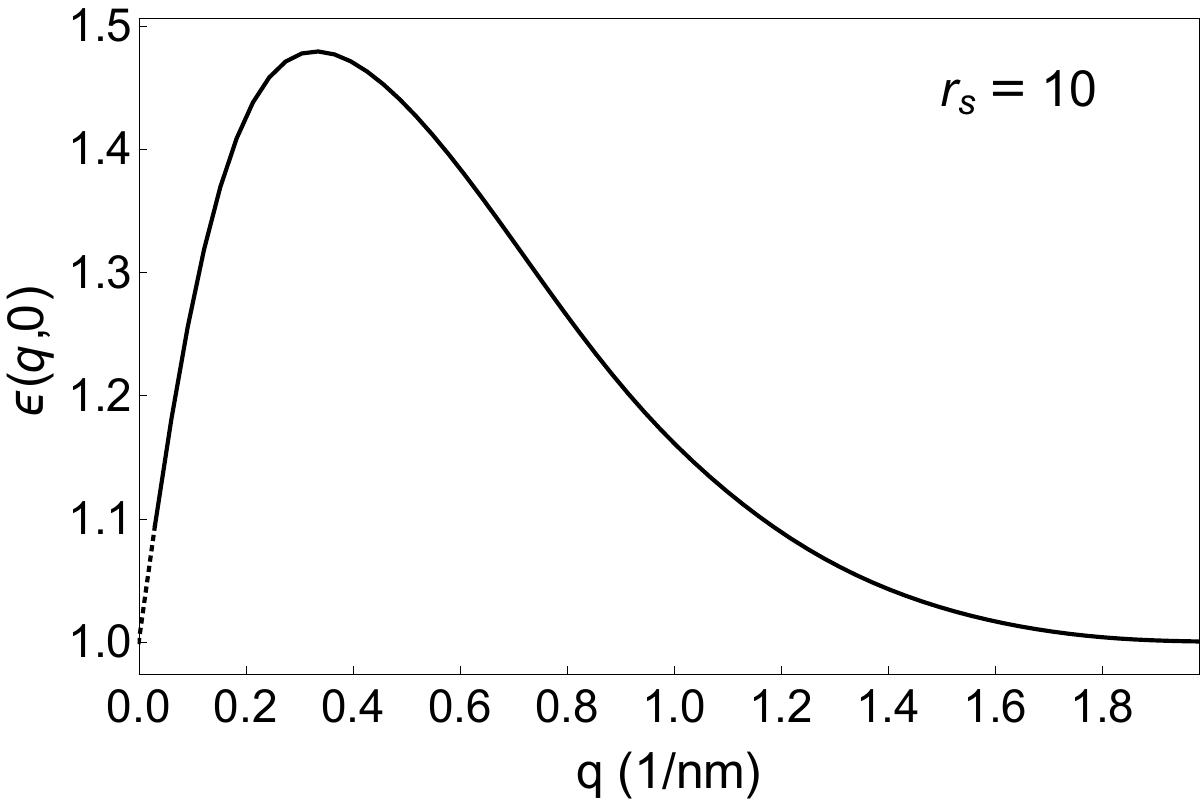}\\[-0.2ex]
\textbf{(h)}
\end{tabular}
\\[0.3ex]

\begin{tabular}{c}
\includegraphics[width=0.238\textwidth]{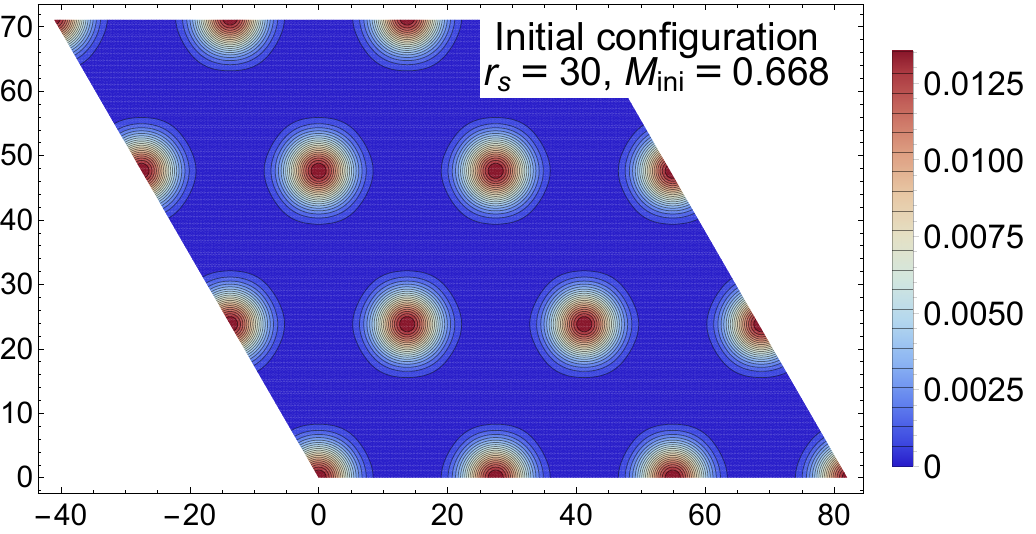}\\[-0.2ex]
\textbf{(i)}
\end{tabular}
&
\begin{tabular}{c}
\includegraphics[width=0.238\textwidth]{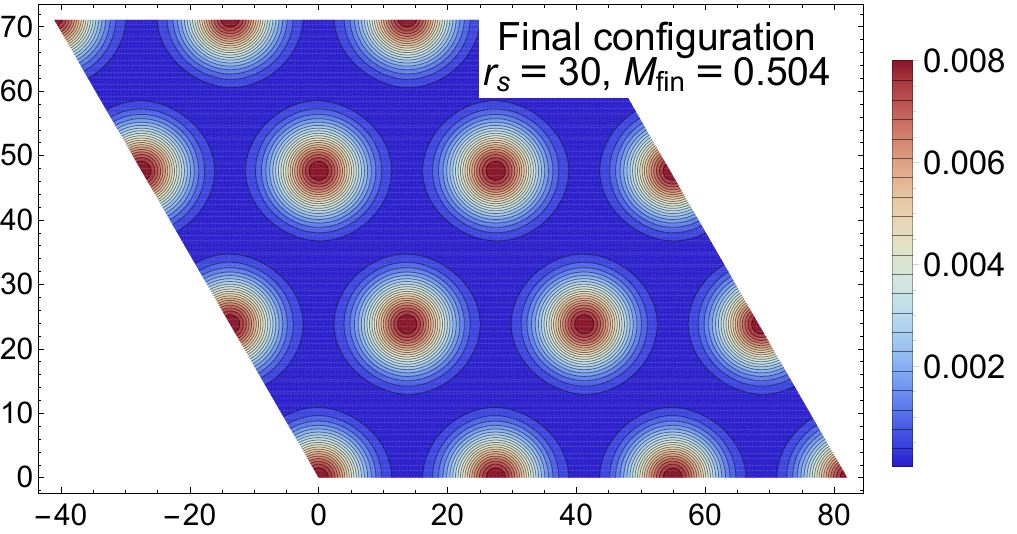}\\[-0.2ex]
\textbf{(j)}
\end{tabular}
&
\begin{tabular}{c}
\includegraphics[width=0.238\textwidth]{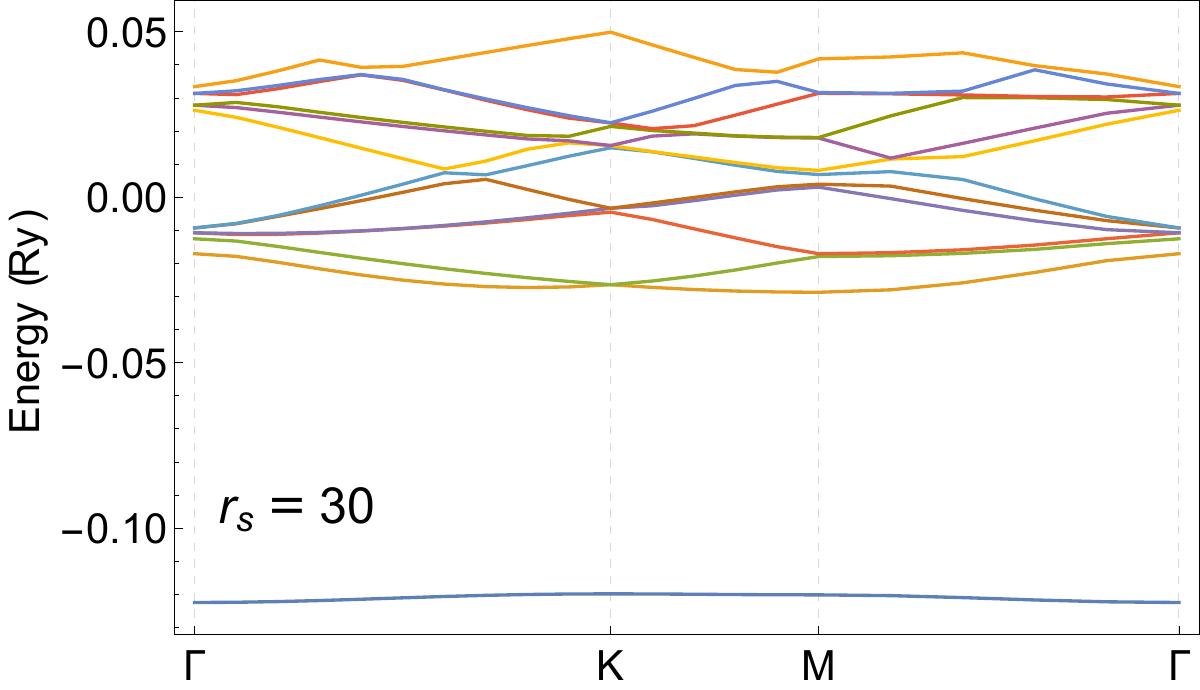}\\[-0.2ex]
\textbf{(k)}
\end{tabular}
&
\begin{tabular}{c}
\includegraphics[width=0.238\textwidth]{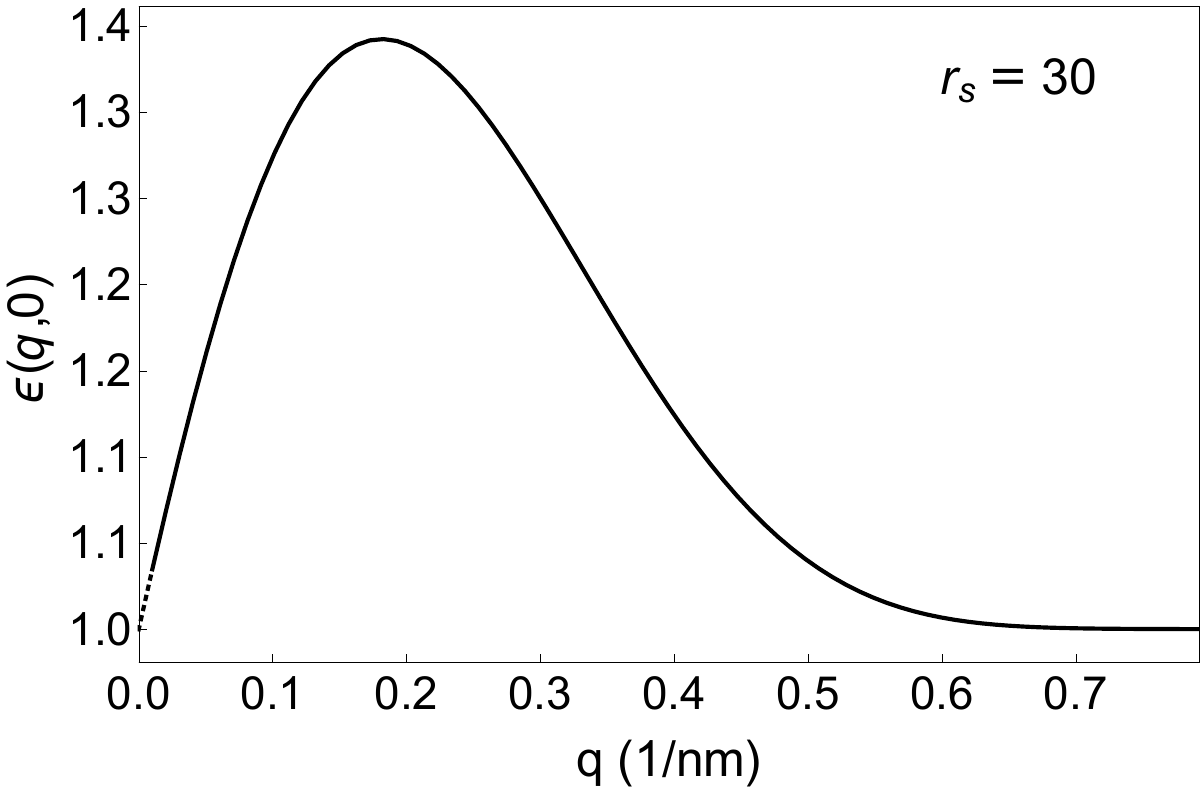}\\[-0.2ex]
\textbf{(l)}
\end{tabular}
\\[0.3ex]

\begin{tabular}{c}
\includegraphics[width=0.238\textwidth]{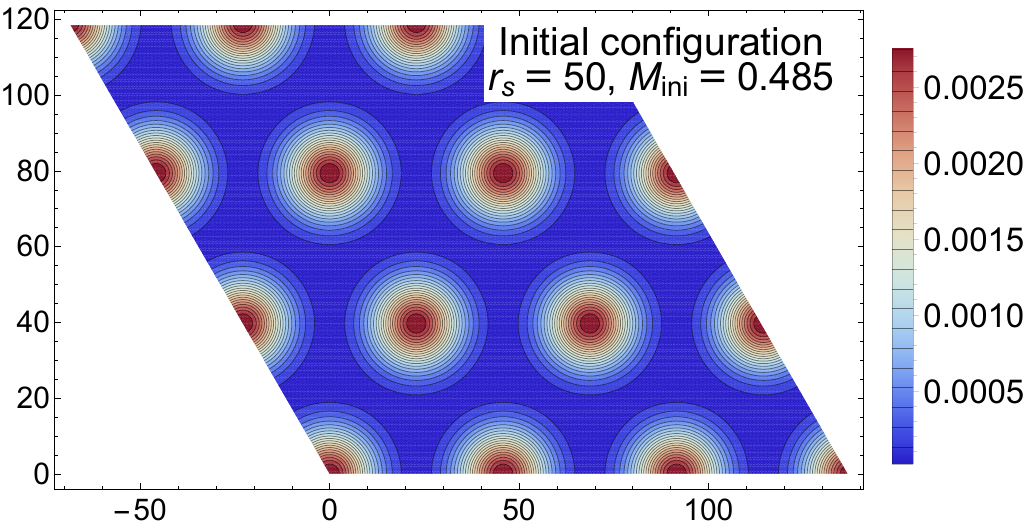}\\[-0.2ex]
\textbf{(m)}
\end{tabular}
&
\begin{tabular}{c}
\includegraphics[width=0.238\textwidth]{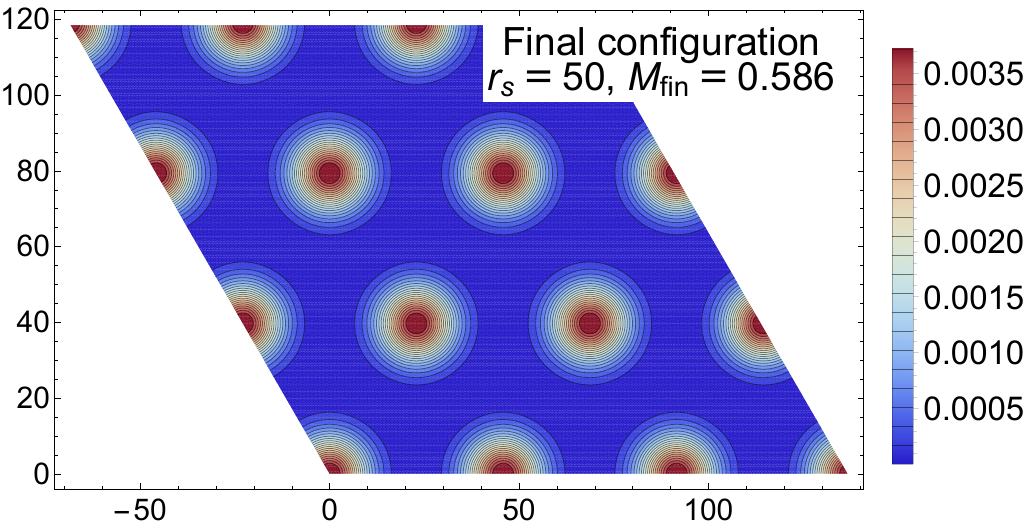}\\[-0.2ex]
\textbf{(n)}
\end{tabular}
&
\begin{tabular}{c}
\includegraphics[width=0.238\textwidth]{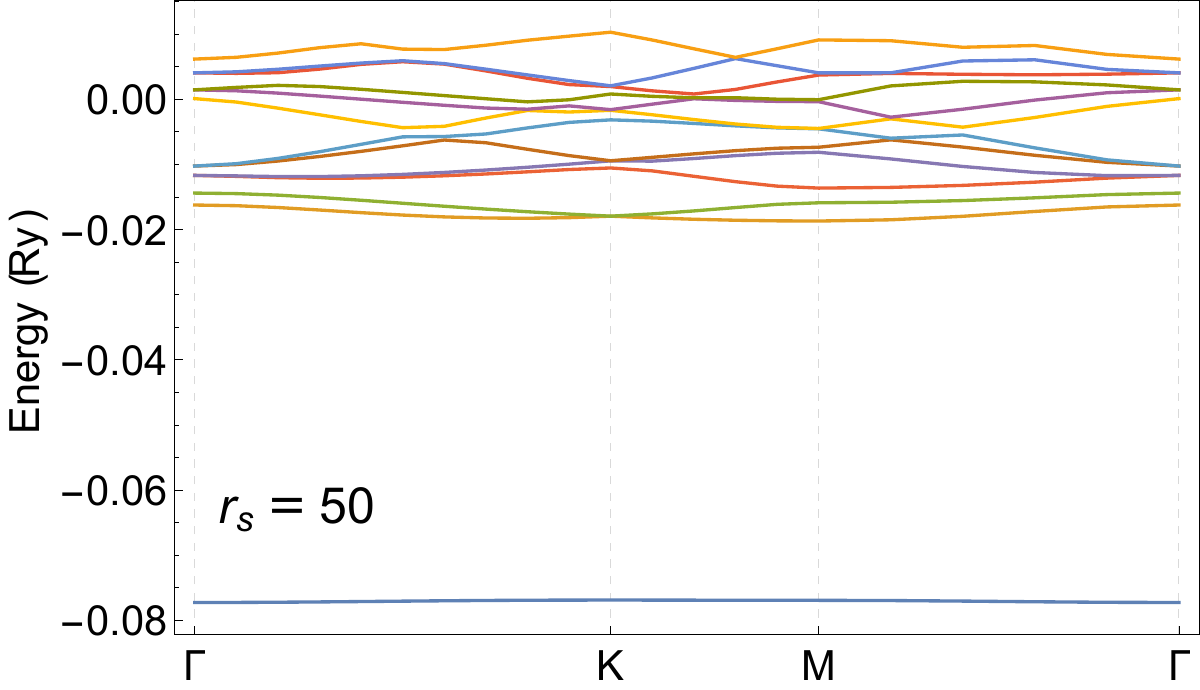}\\[-0.2ex]
\textbf{(o)}
\end{tabular}
&
\begin{tabular}{c}
\includegraphics[width=0.238\textwidth]{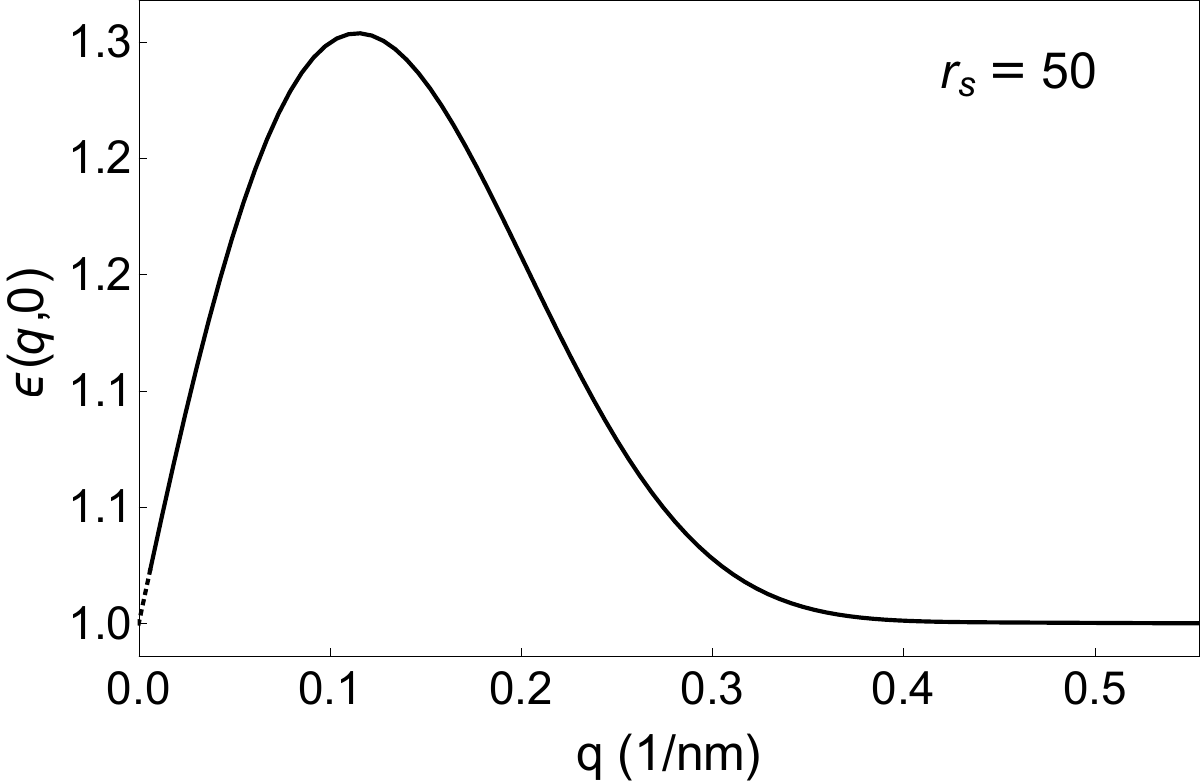}\\[-0.2ex]
\textbf{(p)}
\end{tabular}

\end{tabular}

\caption{
Evolution of the fully spin-polarized Hartree--Fock Wigner-crystal solution with the density parameter $r_s$.
The four rows correspond to $r_s=4$, $10$, $30$, and $50$, as indicated inside the panels.
The columns show, from left to right, the initial charge-density configuration, the converged Hartree--Fock charge density, the same-spin Hartree--Fock band structure along $\Gamma$--$K$--$M$--$\Gamma$, and the static dielectric function $\varepsilon_{\mathrm{HF}}(q,0)$ along $\Gamma$--$K$ direction.
The charge-density panels also indicate the first-star order parameter of the initial and converged configurations, denoted by $M_{\mathrm{ini}}$ and $M_{\mathrm{fin}}$, respectively.
The sequence illustrates the evolution from a weakly crystalline solution at small $r_s$ toward increasingly localized Wigner-crystal states at larger $r_s$, accompanied by changes in the band structure and in the finite-$q$ dielectric response.
The values of $n_{\min}$, $n_{\max}$, and
$C_n=n_{\max}/n_{\min}$ reported in each panel quantify the corresponding
real-space density contrast. For $r_s=4, 10, 30, 50$, the corresponding values of $n_{\min}$, $n_{\max}$,
and $C_n$ are, respectively:
$n_{\min}=5.03\times10^{-2},2.99\times10^{-3},3.46\times10^{-5},2.59\times10^{-6}$,
$n_{\max}=1.38\times10^{-1},4.00\times10^{-2},8.04\times10^{-3},3.72\times10^{-3}$,
and $C_n=2.74,13.35,232,1.44\times10^{3}$.
}
\label{fig:rs_scan_polarized}
\end{figure*}

Figure~\ref{fig:rs_scan_polarized} summarizes a representative fully spin-polarized Hartree--Fock Wigner-crystal solution. The panels show the evolution from the initial crystalline density used to start the self-consistent iteration to the converged charge density, followed by the corresponding same-spin Hartree--Fock band structure and the static dielectric function. This arrangement emphasizes the numerical sequence of the calculation: initialization, self-consistent crystalline solution, quasiparticle spectrum, and density response.

The numerical spectra display the singlet and doublet structures expected from the point-group symmetry of the hexagonal Wigner-crystal Brillouin zone. In particular, the bands remain regular near $\Gamma$, show the appropriate degeneracy structure near the zone corners, and evolve smoothly along the $\Gamma$--$K$--$M$ path. The charge-density and band-structure panels in Fig.~\ref{fig:rs_scan_polarized} are consistent with this diagnostic picture. At larger $r_s$, the converged density is more strongly modulated and the Hartree--Fock spectrum displays clearer crystalline band features. As $r_s$ decreases, the density modulation weakens, the order parameter $M_1$ decreases, and the crystalline Hartree--Fock solution becomes less robust. This behavior indicates a loss of Wigner-crystal character and suggests that the system is approaching the weak-crystal or liquid side, although a thermodynamic identification of the transition would require an energy comparison or an independent phase-stability analysis.


The values of the first-star order parameter included in Fig.~\ref{fig:rs_scan_polarized}
provide a quantitative measure of the density modulation visible in the charge-density
panels. For the converged Hartree--Fock solutions, $M_1^{\mathrm{fin}}$
increases overall with $r_s$, from $M_1^{\mathrm{fin}}=0.100$ at $r_s=4$ to
$M_1^{\mathrm{fin}}=0.586$ at $r_s=50$. This trend confirms that the
self-consistent solution becomes increasingly Wigner-crystal-like in the dilute
regime: as the lattice constant increases and the carrier density decreases,
the charge density becomes more localized around the triangular lattice sites,
and the Fourier weight at the first reciprocal-lattice star increases.

A complementary real-space measure of the density modulation is provided by the
density-contrast ratio
$C_n=n_{\max}/n_{\min}$, where $n_{\max}$ and $n_{\min}$ are the
maximum and minimum values of the converged charge density within one
Wigner-crystal unit cell. While $M_1$ measures the weight of the fundamental
Fourier harmonics, $C_n$ directly quantifies the ripple character of the
real-space density profile. Values of $C_n$ close to unity correspond to a
nearly uniform density, whereas larger values indicate a stronger spatial
separation between density maxima at the lattice sites and density minima in
the interstitial regions. Thus, the combined behavior of $M_1$ and $C_n$
provides a more complete characterization of the Wigner-crystal modulation.

The comparison between the initial and final values also shows that the
self-consistent calculation is not merely preserving the imposed starting
configuration. For example, at $r_s=30$ the initial state has a larger first-star
amplitude, $M_1^{\mathrm{ini}}=0.668$, whereas the converged solution relaxes to
$M_1^{\mathrm{fin}}=0.504$. This reduction is consistent with the visual density
map, where the final density is less sharply localized than the initial
Gaussian configuration. Thus $M_1$ acts as a useful internal consistency check:
the converged value reflects the self-consistent Hartree--Fock balance between
kinetic energy, Hartree repulsion, and exchange, rather than the artificial
degree of localization chosen in the initialization.

\subsection{Static dielectric-function results}
\label{sec:dielectric_results}

The static dielectric functions shown in Fig.~\ref{fig:rs_scan_polarized}
exhibit a systematic reduction of their maximum value with increasing $r_s$.
According to the definition of $\varepsilon_{\mathrm{HF}}(q,0)$ introduced in
Sec.~\ref{sec:static_dielectric_function}, the screened interaction may be
written as
\begin{equation}
\widetilde v_{\mathrm{scr}}(q)
=
\frac{\widetilde v_{\mathrm K}(q)}
{\varepsilon_{\mathrm{HF}}(q,0)} .
\end{equation}
Here $\widetilde v_{\mathrm K}(q)$ is the bare Keldysh interaction, including
the background dielectric environment, whereas
$\varepsilon_{\mathrm{HF}}(q,0)$ describes the additional finite-$q$ screening
generated by the resident carriers in the Hartree--Fock Wigner-crystal state.
Thus larger values of $\varepsilon_{\mathrm{HF}}(q,0)$ correspond to a stronger
reduction of the effective electron--electron interaction at the corresponding
momentum transfer.

The decrease of the peak value of $\varepsilon_{\mathrm{HF}}(q,0)$ with
increasing $r_s$ reflects the reduction of the electronic polarizability in the
dilute Wigner crystal. As $r_s$ increases, the Wigner-crystal lattice constant
increases and the carrier density decreases. The electrons become more
localized, the overlap between neighboring density modulations is reduced, and
the finite-$q$ Hartree--Fock polarizability is weakened. Consequently the
maximum of $\varepsilon_{\mathrm{HF}}(q,0)$ moves closer to unity, indicating
that the resident carriers become less efficient at screening an external
perturbation and that the effective interaction approaches the bare Keldysh
interaction. The momentum at which the maximum occurs gives the most efficient
static-screening momentum channel; its physical interpretation and its scaling
with the Wigner-crystal reciprocal-lattice scale are discussed in SI.6.
\FloatBarrier


\section{Conclusions}
\label{sec:conclusions}

Experimentally isolating spin/valley-polarized Wigner-crystal spectra is
challenging. A possible test of the present predictions for monolayer WSe$_2$
would be a time-resolved, helicity-resolved optical spectroscopy experiment on
an hBN-encapsulated, gate-tunable WSe$_2$ monolayer. In such a protocol, a
resident-electron Wigner crystal in the relevant density range would first be
driven toward a nearly spin/valley-polarized configuration by resonant circular
optical pumping. Efficient optical spin/valley pumping of resident electrons has
already been demonstrated in $n$-doped WSe$_2$ and WS$_2$ monolayers
\cite{Robert2021SpinValleyPumping}, and recent Wigner-polaron experiments
suggest that optical control of spin/valley degrees of freedom in the Wigner
crystal regime is experimentally accessible \cite{Zhang2025WignerPolarons}.
The degree of polarization could be calibrated independently by
Faraday-geometry magneto-optical measurements of the valley-Zeeman splitting in
an out-of-plane magnetic field \cite{Srivastava2015ValleyZeeman}.
Such an experiment, performed on a prepared spin/valley-polarized configuration, would primarily test the spin-conserving, frozen-spin/valley sector of the polarized Hartree--Fock calculation.
The present work connects the conventional theory of the low-density
2D Wigner crystal with recent optical studies of charge order in
atomically thin semiconductors. The triangular Wigner crystal and its collective
properties have long been established as central features of the dilute
2D electron gas \cite{Bonsall1977,Tanatar1989}. Recent experiments
on charge-tunable TMD monolayers have shown that Wigner-crystal charge order can
be accessed optically even in zero magnetic field, through exciton-umklapp
signatures and related resonances \cite{Smolenski2021WignerCrystal}. More recent
work has further emphasized that exciton spectroscopy can probe not only the
static periodic potential of the Wigner crystal, but also dynamical
electron-lattice excitations in the form of Wigner-polaron resonances
\cite{Wang2025WignerCrystalPolarons,Zhang2025WignerPolarons,Adlong2025TheoryExcitonPolarons}.
These developments motivate a microscopic calculation that is adapted to the
Keldysh interaction and optical response of a monolayer TMD.

Within this context, the main physical contribution of the present work is the
self-consistent Hartree--Fock construction of a spin-polarized Wigner-crystal
state in monolayer WSe$_2$, together with the corresponding quasiparticle band
structure and finite-momentum static dielectric response. The calculation
connects, within the same microscopic framework, the real-space crystalline
density, the Hartree--Fock spectrum in the Wigner-crystal Brillouin zone, the
first-star density order parameter $M_1$, the density-contrast ratio $C_n$, and the band-based dielectric
function $\varepsilon_{\mathrm{HF}}(q,0)$. The results show that increasing
$r_s$ strengthens the crystalline density modulation while reducing the
additional electronic screening produced by the resident carriers. This is the
expected behavior of a dilute Wigner crystal: the electrons become more
localized in real space, and their finite-$q$ polarizability becomes less
efficient.

The conclusions should be understood within the scope of the approximation. The
regularity of the band structure, the evolution of $M_1$, and the behavior of
$\varepsilon_{\mathrm{HF}}(q,0)$ are used here as complementary signatures of the
self-consistent crystalline Hartree--Fock solution. They do not by themselves
constitute a thermodynamic proof of the liquid--crystal transition, which would
require an energy comparison or an independent phase-stability analysis.
Nevertheless, the combined density, band-structure, and dielectric-response
diagnostics provide a consistent microscopic picture of the evolution from a
weakly crystalline solution toward a more localized Wigner-crystal state.

The numerical implementation is also flexible enough to support systematic
convergence checks. In particular, the $k$-mesh density can be increased without
changing the structure of the calculation, allowing the high-symmetry band
features and the finite-$q$ dielectric response to be tested on progressively
denser Brillouin-zone grids. The code uses an irreducible
$\Gamma$--$K$--$M$ wedge for the self-consistent calculation and reconstructs the
full Brillouin zone when required for exchange and dielectric matrix elements.
This capability is important for future refinements, including denser momentum
sampling, larger plane-wave bases, alternative interaction kernels, and
extensions toward unpolarized, multiband, finite-temperature, or collective-mode
calculations.

The numerical code used in this work is a research version developed for the present calculations; it is not yet distributed as a public software package, but can be made available from the authors upon reasonable request.

\section*{Acknowledgements}

Y.C. Chang acknowledges support from National Sci-
ence and Technology Council, Taiwan under grant nos.
NSTC 114-2221-E-001-001-MY2 and NSTC 114-2112-M-006-030''.

\printbibliography

\clearpage
\vspace*{1em}
\begin{center}
{\LARGE\bfseries Supporting Information}
\end{center}
\vspace{1em}

\addcontentsline{toc}{section}{Supporting Information}

\setcounter{section}{0}
\setcounter{subsection}{0}
\setcounter{equation}{0}
\setcounter{figure}{0}
\setcounter{table}{0}

\renewcommand{\thesection}{SI.\arabic{section}}
\renewcommand{\thesubsection}{\thesection.\arabic{subsection}}
\renewcommand{\theequation}{S\arabic{equation}}
\renewcommand{\thefigure}{S\arabic{figure}}
\renewcommand{\thetable}{S\arabic{table}}

\addcontentsline{toc}{section}{Supporting Information}

\section{\texorpdfstring{Hartree--Fock matrix elements in the $\mathbf{k}+\mathbf{G}$ representation}{Hartree--Fock matrix elements in the k+G representation}}
\label{sec:si_hf_matrix_elements}

This section gives the coordinate-space starting point and the explicit
projection leading to the matrix elements used in Sec.~\ref{sec:hf_kG}. The Hartree--Fock equation in coordinate representation is
\begin{equation}
\int d^2r'\,
H_{\mathrm{HF}}(\mathbf r,\mathbf r')
\psi_{n\mathbf{k}\sigma}(\mathbf r')
=
\varepsilon_{n\mathbf{k}\sigma}
\psi_{n\mathbf{k}\sigma}(\mathbf r),
\label{eq:si_hf_real}
\end{equation}
where, in the absence of an external potential,
\begin{equation}
H_{\mathrm{HF}}(\mathbf r,\mathbf r')
=
\left[
h_{\mathrm{band}}(\mathbf r)+V_H(\mathbf r)
\right]
\delta(\mathbf r-\mathbf r')
-
\Sigma_F^\sigma(\mathbf r,\mathbf r') .
\label{eq:si_hf_kernel}
\end{equation}
Here $h_{\mathrm{band}}$ is the effective band-edge operator, $V_H$ is the Hartree potential, and $\Sigma_F^\sigma$ is the spin-dependent Fock exchange operator. The local Hartree potential is
\begin{equation}
V_H(\mathbf r)
=
\int d^2r'\,
v(\mathbf r-\mathbf r')
n(\mathbf r'),
\label{eq:si_hartree_potential}
\end{equation}
whereas the exchange kernel is
\begin{equation}
\Sigma_F^\sigma(\mathbf r,\mathbf r')
=
v(\mathbf r-\mathbf r')
\rho_\sigma(\mathbf r,\mathbf r').
\label{eq:si_fock_kernel}
\end{equation}
The spin-resolved one-body density matrix is
\begin{equation}
\rho_\sigma(\mathbf r,\mathbf r')
=
\sum_{n\in\mathrm{occ}}
\sum_{\mathbf k}
w_{\mathbf k}
\psi_{n\mathbf{k}\sigma}(\mathbf r)
\psi^*_{n\mathbf{k}\sigma}(\mathbf r') ,
\label{eq:si_density_matrix}
\end{equation}
where $w_{\mathbf k}$ are the Brillouin-zone integration weights.

Since the Wigner-crystal state is periodic with respect to the emergent triangular lattice, the single-particle orbitals are expanded as
\begin{equation}
\psi_{n\mathbf{k}\sigma}(\mathbf r)
=
\sum_{\mathbf G}
Z_{n\mathbf G\sigma}(\mathbf k)
\frac{1}{\sqrt{A}}
e^{i(\mathbf k+\mathbf G)\cdot\mathbf r},
\label{eq:si_bloch_expansion}
\end{equation}
where $A=N_cA_W$ is the total normalization area, $A_W$ is the real-space Wigner-crystal primitive-cell area, and $N_c$ is the number of Wigner cells in the Born--von Karman supercell. Equivalently,
\begin{equation}
\left|\psi_{n\mathbf{k}\sigma}\right\rangle
=
\sum_{\mathbf G}
Z_{n\mathbf G\sigma}(\mathbf k)
\left|\mathbf k+\mathbf G\right\rangle .
\label{eq:si_bloch_vector}
\end{equation}
Projecting Eq.~\eqref{eq:si_hf_real} onto $\left\langle \mathbf k+\mathbf G\right|$ gives
\begin{equation}
\sum_{\mathbf G'}
\left\langle
\mathbf k+\mathbf G
\left|
H_{\mathrm{HF}}
\right|
\mathbf k+\mathbf G'
\right\rangle
Z_{n\mathbf G'\sigma}(\mathbf k)
=
\varepsilon_{n\mathbf{k}\sigma}
Z_{n\mathbf G\sigma}(\mathbf k).
\label{eq:si_projected_hf}
\end{equation}
Defining
\begin{equation}
h_{\mathbf G\mathbf G'}^\sigma(\mathbf k)
=
\left\langle
\mathbf k+\mathbf G
\left|
H_{\mathrm{HF}}
\right|
\mathbf k+\mathbf G'
\right\rangle ,
\label{eq:si_hmatrix_def}
\end{equation}
one obtains the matrix eigenvalue problem
\begin{equation}
\sum_{\mathbf G'}
h_{\mathbf G\mathbf G'}^\sigma(\mathbf k)
Z_{n\mathbf G'\sigma}(\mathbf k)
=
\varepsilon_{n\mathbf{k}\sigma}
Z_{n\mathbf G\sigma}(\mathbf k).
\label{eq:si_hf_matrix}
\end{equation}
For a parabolic effective-mass band,
\begin{equation}
h_{\mathrm{band}}
=
-\frac{\hbar^2}{2m^*}\nabla^2 .
\label{eq:si_band_operator}
\end{equation}
The plane waves are eigenstates of this operator. Therefore, the kinetic contribution is diagonal in the $\mathbf G$ index,
\begin{equation}
T_{\mathbf G\mathbf G'}(\mathbf k)
=
\frac{\hbar^2|\mathbf k+\mathbf G|^2}{2m^*}
\delta_{\mathbf G\mathbf G'} .
\label{eq:si_kinetic_matrix}
\end{equation}
The Hartree potential is periodic with the Wigner-crystal lattice and can be expanded as
\begin{equation}
V_H(\mathbf r)
=
\sum_{\mathbf Q}
V_H(\mathbf Q)e^{i\mathbf Q\cdot\mathbf r}.
\label{eq:si_hartree_fourier}
\end{equation}
Using the normalized plane-wave representation,
\begin{equation}
\left\langle \mathbf r \middle| \mathbf k+\mathbf G \right\rangle
=
\frac{1}{\sqrt{A}}
e^{i(\mathbf k+\mathbf G)\cdot\mathbf r},
\label{eq:si_plane_wave}
\end{equation}
the Hartree matrix element is
\begin{align}
V^H_{\mathbf G\mathbf G'}
&=
\left\langle
\mathbf k+\mathbf G
\left|
V_H
\right|
\mathbf k+\mathbf G'
\right\rangle
\nonumber\\
&=
\sum_{\mathbf Q}V_H(\mathbf Q)
\frac{1}{A}
\int_A d^2r\,
e^{i(\mathbf Q+\mathbf G'-\mathbf G)\cdot\mathbf r}.
\label{eq:si_hartree_projection}
\end{align}
The integral gives the reciprocal-lattice selection rule
\begin{equation}
\frac{1}{A}
\int_A d^2r\,
e^{i(\mathbf Q+\mathbf G'-\mathbf G)\cdot\mathbf r}
=
\delta_{\mathbf Q,\mathbf G-\mathbf G'}.
\label{eq:si_hartree_selection}
\end{equation}
Therefore
\begin{equation}
V^H_{\mathbf G\mathbf G'}
=
V_H(\mathbf G-\mathbf G').
\label{eq:si_hartree_matrix_1}
\end{equation}
The Hartree potential is the convolution of the interaction with the density. In reciprocal space this gives
\begin{equation}
V_H(\mathbf Q)
=
\widetilde{v}(\mathbf Q)n(\mathbf Q),
\label{eq:si_hartree_convolution}
\end{equation}
and hence
\begin{equation}
V^H_{\mathbf G\mathbf G'}
=
\widetilde{v}(\mathbf G-\mathbf G')
n(\mathbf G-\mathbf G').
\label{eq:si_hartree_matrix_2}
\end{equation}
This contribution is independent of $\mathbf k$.
The Fock matrix element is
\begin{equation}
\Sigma_{\mathbf G\mathbf G'}^{F,\sigma}(\mathbf k)
=
\left\langle
\mathbf k+\mathbf G
\left|
\Sigma_F^\sigma
\right|
\mathbf k+\mathbf G'
\right\rangle .
\label{eq:si_fock_def}
\end{equation}
Substituting Eqs.~\eqref{eq:si_fock_kernel} and \eqref{eq:si_density_matrix}, and using the expansion in Eq.~\eqref{eq:si_bloch_expansion}, gives
\begin{align}
\Sigma_{\mathbf G\mathbf G'}^{F,\sigma}(\mathbf k)
&=
\sum_{n'\in\mathrm{occ}}
\sum_{\mathbf k'}
w_{\mathbf k'}
\sum_{\mathbf G_1,\mathbf G_2}
Z_{n'\mathbf G_1\sigma}(\mathbf k')
Z^*_{n'\mathbf G_2\sigma}(\mathbf k')
\nonumber\\
&\quad\times
I(\mathbf G,\mathbf G';\mathbf k,\mathbf k';\mathbf G_1,\mathbf G_2),
\label{eq:si_fock_intermediate}
\end{align}
where
\begin{align}
I
&=
\int d^2r\,d^2r'\,
v(\mathbf r-\mathbf r')
e^{i[(\mathbf k'+\mathbf G_1)-(\mathbf k+\mathbf G)]\cdot\mathbf r}
e^{i[(\mathbf k+\mathbf G')-(\mathbf k'+\mathbf G_2)]\cdot\mathbf r'} .
\label{eq:si_I_def}
\end{align}
The interaction is Fourier transformed as
\begin{equation}
v(\mathbf r-\mathbf r')
=
\int\frac{d^2q}{(2\pi)^2}
\widetilde{v}(\mathbf q)
e^{i\mathbf q\cdot(\mathbf r-\mathbf r')}.
\label{eq:si_interaction_transform}
\end{equation}
The integrations over $\mathbf r$ and $\mathbf r'$ impose
\begin{equation}
\mathbf q
=
(\mathbf k+\mathbf G)-(\mathbf k'+\mathbf G_1),
\qquad
\mathbf G-\mathbf G'
=
\mathbf G_1-\mathbf G_2 .
\label{eq:si_fock_selection}
\end{equation}
Thus
\begin{equation}
I
=
\widetilde{v}
\!\left[
(\mathbf k+\mathbf G)-(\mathbf k'+\mathbf G_1)
\right]
\delta_{\mathbf G-\mathbf G',\,\mathbf G_1-\mathbf G_2}.
\label{eq:si_I_result}
\end{equation}
Using the selection rule to eliminate $\mathbf G_2$, the exchange matrix becomes
\begin{align}
\Sigma_{\mathbf G\mathbf G'}^{F,\sigma}(\mathbf k)
&=
\sum_{n'\in\mathrm{occ}}
\sum_{\mathbf k'}
w_{\mathbf k'}
\sum_{\mathbf G_1}
Z_{n'\mathbf G_1\sigma}(\mathbf k')
Z^*_{n',\mathbf G_1-\mathbf G+\mathbf G',\sigma}(\mathbf k')
\nonumber\\
&\quad\times
\widetilde{v}
\!\left[
(\mathbf k+\mathbf G)-(\mathbf k'+\mathbf G_1)
\right].
\label{eq:si_fock_final}
\end{align}
Terms for which the shifted reciprocal vector
$\mathbf G_1-\mathbf G+\mathbf G'$ lies outside the retained plane-wave basis are omitted.

Combining the kinetic, Hartree, and Fock terms gives
\begin{equation}
h_{\mathbf G\mathbf G'}^\sigma(\mathbf k)
=
T_{\mathbf G\mathbf G'}(\mathbf k)
+
V^H_{\mathbf G\mathbf G'}
-
\Sigma_{\mathbf G\mathbf G'}^{F,\sigma}(\mathbf k).
\label{eq:si_hf_final}
\end{equation}
For the spin-polarized Wigner crystal considered in the main text, only one spin sector is occupied, and the exchange contribution is present only in the same-spin particle channel.

\section{\texorpdfstring{Initial density harmonics $\eta_W^{(0)}(\mathbf Q)$}{Initial density harmonics etaW0(Q)}}
\label{Ini_dens}

The relation between the initial coefficients $F_{\mathbf k}^{(0)}(\mathbf G)$ and the initial density harmonics $\eta_W^{(0)}(\mathbf Q)$ follows directly from the density associated with the cell-periodic part of the Bloch state. Since the overall phase factor $e^{i\mathbf k\cdot\mathbf r}$ cancels in the density, we define
\begin{equation}
u_{\mathbf k}^{(0)}(\mathbf r)
=
\sum_{\mathbf G}
F_{\mathbf k}^{(0)}(\mathbf G)
e^{i\mathbf G\cdot\mathbf r}.
\label{eq:si_initial_u}
\end{equation}
The initial density is written as the Brillouin-zone average
\begin{equation}
n^{(0)}(\mathbf r)
=
\frac{1}{A_W}
\sum_{\mathbf k}
w_{\mathbf k}
\left|
u_{\mathbf k}^{(0)}(\mathbf r)
\right|^2 ,
\label{eq:si_initial_density_average}
\end{equation}
with $\sum_{\mathbf k}w_{\mathbf k}=1$. Substituting the plane-wave expansion of $u_{\mathbf k}^{(0)}(\mathbf r)$ gives
\begin{align}
n^{(0)}(\mathbf r)
&=
\frac{1}{A_W}
\sum_{\mathbf k}
w_{\mathbf k}
\sum_{\mathbf G,\mathbf G'}
F_{\mathbf k}^{(0)}(\mathbf G)
F_{\mathbf k}^{(0)*}(\mathbf G')
e^{i(\mathbf G-\mathbf G')\cdot\mathbf r}.
\label{eq:si_initial_density_expand}
\end{align}
Comparing this expression with
\begin{equation}
n^{(0)}(\mathbf r)
=
\sum_{\mathbf Q}
\eta_W^{(0)}(\mathbf Q)
e^{i\mathbf Q\cdot\mathbf r},
\label{eq:si_initial_density_fourier}
\end{equation}
one obtains
\begin{equation}
\eta_W^{(0)}(\mathbf Q)
=
\frac{1}{A_W}
\sum_{\mathbf k}
w_{\mathbf k}
\sum_{\mathbf G}
F_{\mathbf k}^{(0)*}(\mathbf G)
F_{\mathbf k}^{(0)}(\mathbf G+\mathbf Q).
\label{eq:si_initial_etaW_general}
\end{equation}
For the real Gaussian initialization used in the code, this becomes
\begin{equation}
\eta_W^{(0)}(\mathbf Q)
=
\frac{1}{A_W}
\sum_{\mathbf k}
w_{\mathbf k}
\sum_{\mathbf G}
F_{\mathbf k}^{(0)}(\mathbf G)
F_{\mathbf k}^{(0)}(\mathbf G+\mathbf Q).
\label{eq:si_initial_etaW_real}
\end{equation}
For a uniform full-BZ mesh, $w_{\mathbf k}=1/N_k$, so Eq.~\eqref{eq:si_initial_etaW_real} is identical to the explicit $1/(A_WN_k)$ expression.


\section{Static Hartree--Fock band polarizability}
\label{sec:si_static_polarizability}

The density response is computed from the converged Hartree--Fock eigenvalues and eigenvectors in the $\mathbf k+\mathbf G$ basis. The Bloch states are
\begin{equation}
\psi_{n\mathbf k}(\mathbf r)
=
\frac{1}{\sqrt A}
\sum_{\mathbf G}
Z_{n\mathbf G}(\mathbf k)
e^{i(\mathbf k+\mathbf G)\cdot\mathbf r}.
\label{eq:si_bloch_for_response}
\end{equation}
For a perturbing wave vector $\mathbf q$, the final momentum is folded back into the first Wigner-crystal Brillouin zone,
\begin{equation}
\mathbf k_{\mathbf q}
=
\mathbf k+\mathbf q-\mathbf G_{\mathrm f}(\mathbf k,\mathbf q),
\qquad
\mathbf k_{\mathbf q}\in \mathrm{1BZ},
\label{eq:si_kq_fold}
\end{equation}
where $\mathbf G_{\mathrm f}$ is a reciprocal-lattice vector. The density matrix element is
\begin{align}
M_{nm}(\mathbf k,\mathbf q)
&=
\left\langle
n\mathbf k
\left|
e^{-i\mathbf q\cdot\mathbf r}
\right|
m\mathbf k_{\mathbf q}
\right\rangle
\nonumber\\
&=
\sum_{\mathbf G,\mathbf G'}
Z^*_{n\mathbf G}(\mathbf k)
Z_{m\mathbf G'}(\mathbf k_{\mathbf q})
\frac{1}{A}
\int_A d^2r\,
e^{-i(\mathbf k+\mathbf G)\cdot\mathbf r}
e^{-i\mathbf q\cdot\mathbf r}
e^{i(\mathbf k_{\mathbf q}+\mathbf G')\cdot\mathbf r}.
\label{eq:si_M_start}
\end{align}
Using Eq.~\eqref{eq:si_kq_fold}, the phase in the integral is
\begin{equation}
\mathbf k_{\mathbf q}+\mathbf G'
-\mathbf q-\mathbf k-\mathbf G
=
\mathbf G'-\mathbf G-\mathbf G_{\mathrm f}.
\label{eq:si_M_selection_phase}
\end{equation}
Therefore the plane-wave integral imposes the reciprocal-lattice selection rule
\begin{equation}
\mathbf G'=\mathbf G+\mathbf G_{\mathrm f},
\label{eq:si_M_selection}
\end{equation}
and the form factor becomes
\begin{equation}
M_{nm}(\mathbf k,\mathbf q)
=
\sum_{\mathbf G}
Z^*_{n\mathbf G}(\mathbf k)
Z_{m,\mathbf G+\mathbf G_{\mathrm f}}(\mathbf k_{\mathbf q}) .
\label{eq:si_M_final}
\end{equation}
Terms for which $\mathbf G+\mathbf G_{\mathrm f}$ is outside the retained plane-wave basis are omitted in the numerical implementation.

With the convention used in the main text, the positive static polarizability is the negative of the usual static density-response function. For a fully spin-polarized insulator at zero temperature it can be written as an occupied--empty interband sum,
\begin{equation}
\Pi_{\mathrm{HF}}(\mathbf q,0)
=
\frac{1}{A_W}
\sum_{\mathbf k}
w_{\mathbf k}
\sum_{n\in\mathrm{occ}}
\sum_{m\in\mathrm{emp}}
\left|
M_{nm}(\mathbf k,\mathbf q)
\right|^2
\frac{
\Delta E_{mn}(\mathbf k,\mathbf q)
}{
\Delta E_{mn}^2(\mathbf k,\mathbf q)+\eta_{\Pi}^2
},
\label{eq:si_PiHF_static}
\end{equation}
where
\begin{equation}
\Delta E_{mn}(\mathbf k,\mathbf q)
=
\varepsilon_{m\mathbf k_{\mathbf q}}
-
\varepsilon_{n\mathbf k}.
\label{eq:si_DeltaE}
\end{equation}
The parameter $\eta_{\Pi}$ is a small numerical broadening. In the analytic static limit, $\eta_{\Pi}\rightarrow0^+$ and the factor in Eq.~\eqref{eq:si_PiHF_static} becomes $1/\Delta E_{mn}$.

\subsection{Long-wavelength limit of the dielectric function}
\label{sec:si_dielectric_long_wavelength}

At $\mathbf q=\mathbf0$, one has $\mathbf G_{\mathrm f}=\mathbf0$ and $\mathbf k_{\mathbf q}=\mathbf k$. Equation~\eqref{eq:si_M_final} gives
\begin{equation}
M_{nm}(\mathbf k,\mathbf0)
=
\sum_{\mathbf G}
Z^*_{n\mathbf G}(\mathbf k)
Z_{m\mathbf G}(\mathbf k)
=
\delta_{nm},
\label{eq:si_M_q0}
\end{equation}
by orthonormality of the Hartree--Fock eigenvectors at fixed $\mathbf k$. Since the polarizability in Eq.~\eqref{eq:si_PiHF_static} contains only occupied--empty terms, $n\in\mathrm{occ}$ and $m\in\mathrm{emp}$, Eq.~\eqref{eq:si_M_q0} implies
\begin{equation}
\Pi_{\mathrm{HF}}(\mathbf0,0)=0 .
\label{eq:si_Pi_q0_zero}
\end{equation}

For sufficiently small nonzero $\mathbf q$, no reciprocal-lattice folding is required for generic $\mathbf k$, and the matrix element can be written as an overlap between cell-periodic functions,
\begin{equation}
M_{nm}(\mathbf k,\mathbf q)
=
\left\langle
u_{n\mathbf k}
\middle|
u_{m,\mathbf k+\mathbf q}
\right\rangle .
\label{eq:si_M_cell_periodic}
\end{equation}
Expanding the final state in powers of $\mathbf q$ gives
\begin{equation}
u_{m,\mathbf k+\mathbf q}
=
u_{m\mathbf k}
+
\sum_a q_a\,\partial_{k_a}u_{m\mathbf k}
+
O(q^2),
\label{eq:si_u_expand}
\end{equation}
and therefore
\begin{equation}
M_{nm}(\mathbf k,\mathbf q)
=
\delta_{nm}
+
\sum_a q_a
\left\langle
u_{n\mathbf k}
\middle|
\partial_{k_a}u_{m\mathbf k}
\right\rangle
+
O(q^2).
\label{eq:si_M_smallq}
\end{equation}
For an interband occupied--empty transition, $n\ne m$, the zeroth-order term vanishes and
\begin{equation}
M_{nm}(\mathbf k,\mathbf q)=O(q),
\qquad
\left|M_{nm}(\mathbf k,\mathbf q)\right|^2=O(q^2).
\label{eq:si_M_interband_q}
\end{equation}
Because the Wigner-crystal state is insulating, the occupied and empty Hartree--Fock bands are separated by a finite gap,
\begin{equation}
\Delta E_{mn}(\mathbf k,\mathbf q)\geq \Delta_{\mathrm{HF}}>0
\label{eq:si_gap_condition}
\end{equation}
for sufficiently small $\mathbf q$. Combining Eqs.~\eqref{eq:si_PiHF_static} and \eqref{eq:si_M_interband_q} gives
\begin{equation}
\Pi_{\mathrm{HF}}(\mathbf q,0)=O(q^2),
\qquad
q\rightarrow0 .
\label{eq:si_Pi_smallq}
\end{equation}

The static dielectric function is
\begin{equation}
\varepsilon_{\mathrm{HF}}(\mathbf q,0)
=
1+
\widetilde v_{\mathrm K}(q)
\Pi_{\mathrm{HF}}(\mathbf q,0).
\label{eq:si_epsilon_def}
\end{equation}
For the two-dimensional Keldysh interaction,
\begin{equation}
\widetilde v_{\mathrm K}(q)
=
\frac{2\pi e^2}{\varepsilon q(1+q\rho_0)}
=
\frac{2\pi e^2}{\varepsilon q}
+
O(1)
\qquad
(q\rightarrow0).
\label{eq:si_keldysh_smallq}
\end{equation}
Therefore
\begin{equation}
\widetilde v_{\mathrm K}(q)
\Pi_{\mathrm{HF}}(\mathbf q,0)
=
O(q),
\qquad
q\rightarrow0,
\label{eq:si_product_smallq}
\end{equation}
and the dielectric function is continuous at the origin:
\begin{equation}
\varepsilon_{\mathrm{HF}}(\mathbf0,0)=1,
\qquad
\lim_{q\rightarrow0}
\varepsilon_{\mathrm{HF}}(\mathbf q,0)=1 .
\label{eq:si_epsilon_continuity}
\end{equation}
This result is a consequence of the finite insulating gap and the orthogonality of the Hartree--Fock Bloch eigenvectors.

\section{\texorpdfstring{Reciprocal-space basis and numerical parameters}{Reciprocal-space basis and numerical parameters}}
\label{sec:si_miller_map}
The twelve symmetry operations consist of the six rotations $C_6^j$, $j=0,\ldots,5$, and six mirror operations $\sigma_j$. In Miller indices they are represented as
\begin{equation}
\begin{array}{c|c|c}
\alpha & \mathrm{operation} & (m,n)\mapsto(m_\alpha,n_\alpha)\\
\hline
1  & E   & (m,n)\\
2  & C_6^1   & (-m-n,m)\\
3  & C_6^2   & (n,-m-n)\\
4  & C_6^3   & (-m,-n)\\
5  & C_6^4   & (m+n,-m)\\
6  & C_6^5   & (-n,m+n)\\
7  & \sigma_1 & (-m-n,n)\\
8  & \sigma_2 & (n,m)\\
9  & \sigma_3 & (m,-m-n)\\
10 & \sigma_4 & (m+n,-n)\\
11 & \sigma_5 & (-n,-m)\\
12 & \sigma_6 & (-m,m+n)
\end{array}
\label{eq:D6_miller_operations}
\end{equation}
The twelve operations consist of the six rotations $C_6^j$, $j=0,\ldots,5$, and six mirror operations $\sigma_j$. The map in Miller indices
is used to generate symmetry images of the k + G expansion while preserving |G| and the closure of the truncated basis.
The numerical basis stores the index map
$\mathbf G_{mn}\mapsto \mathbf G_{m_\alpha n_\alpha}$ for all retained vectors and all twelve operations. This map is used to generate symmetry images of the $\mathbf k+\mathbf G$ expansion while preserving $|\mathbf G|$ and the closure of the truncated basis.

In reduced reciprocal coordinates $\mathbf k=x\mathbf b_1+y\mathbf b_2$ and the wedge vertices are $\Gamma=(0,0), K=(1/3, 1/3),  M=(0,1/2)$.
For a wedge resolution $N_{\mathrm{side}}$, the mesh is generated row by row. The row index is $j=0,\ldots,N_{\mathrm{side}}$, and the point index in that row is
$t=0,\ldots,2(N_{\mathrm{side}}-j)$. The corresponding reduced coordinates are
\begin{equation}
\mathbf k_{jt}
=
\frac{t}{6N_{\mathrm{side}}}\mathbf b_1
+
\frac{3j+t}{6N_{\mathrm{side}}}\mathbf b_2 ,
\label{eq:wedge_kmesh}
\end{equation}
and the number of representative wedge points is
$N_w=\sum_{j=0}^{N_{\mathrm{side}}}\left[2(N_{\mathrm{side}}-j)+1\right]=(N_{\mathrm{side}}+1)^2$.

Each wedge point represents a set of symmetry-related momenta in the full hexagonal Brillouin zone. Its multiplicity is the number of distinct images generated by the twelve $D_6$ operations, after identifying momenta equivalent on the periodic Brillouin-zone torus. In the present mesh this gives multiplicity one for $\Gamma$, two for $K$, three for $M$, six for non-vertex wedge-boundary points, and twelve for interior wedge points. These multiplicities define the Brillouin-zone weights used in the self-consistent Hartree--Fock sums.

The numerical parameters used in the Hartree--Fock calculations are summarized in Table~\ref{tab:numerical_parameters}.

\begin{table}[t]
\centering
\caption{Representative numerical parameters used in the Hartree--Fock and dielectric-function calculations.}
\label{tab:numerical_parameters}
\begin{tabular}{ll}
\hline
Parameter & Value / meaning \\
\hline
$N_{\mathrm{side}}$ & 5-20 / wedge-mesh resolution \\
$N_w=(N_{\mathrm{side}}+1)^2$ & 36-441  / number of irreducible wedge points \\
$N_{\mathrm{full}}$ & 300-4800 number of reconstructed full-BZ points \\
$N_G$ & 50-120 / number of retained reciprocal vectors \\
$N_b$ & 13-17 / number of Hartree--Fock bands retained \\
$N_{\mathrm{occ}}$ &  $N_{\mathrm{occ}}=1$ in the spin-polarized calculation \\
$\eta_{\Pi}$ &   $5.0\times10^{-4}~\mathrm{Ry}^{*}$ in the static polarizability \\
$\rho_0$ & $1.12~\mathrm{nm}$ / Keldysh screening length \\
$m^*$ & $0.4$ / effective electron mass in WSe$_2$ \\
$\varepsilon$ & $4$ / environmental dielectric constant \\
\hline
\end{tabular}
\end{table}

\section{Full-zone exchange matrix used for the band-structure calculation}
\label{sec:Full_zone}
The production Hartree--Fock calculation evaluates the exchange matrix by an explicit summation over the full Brillouin-zone mesh. This full mesh is generated from the representative points in the irreducible $\Gamma$--$K$--$M$ wedge. If $\mathbf k_w$ is a wedge representative and $s_w$ is its symmetry multiplicity, the distinct full-zone images are
\begin{equation}
\mathbf k_{w j}
=
S_{\alpha_j(w)}\mathbf k_w,
\qquad
j=1,\ldots,s_w ,
\label{eq:si_full_bz_images}
\end{equation}
where $S_{\alpha}$ denotes one of the twelve $D_6$ operations and $\alpha_j(w)$ is the compact list of operations that generate distinct images of $\mathbf k_w$ on the periodic Brillouin-zone torus. The full-zone mesh size is therefore
\begin{equation}
N_{\mathrm{full}}
=
\sum_w s_w ,
\label{eq:si_full_bz_size}
\end{equation}
and the weights satisfy $\sum_{\ell=1}^{N_{\mathrm{full}}}w_\ell=1$. For a uniform full-zone mesh, $w_\ell=1/N_{\mathrm{full}}$.

The eigenvectors on the full zone are reconstructed from the wedge eigenvectors using the reciprocal-vector symmetry map. If $\mathbf k_\ell=S_\alpha\mathbf k_w$, then
\begin{equation}
Z^{\mathrm{full}}_{\nu\mathbf G}(\mathbf k_\ell)
=
Z^{\mathrm{wedge}}_{\nu,S_\alpha^{-1}\mathbf G}(\mathbf k_w).
\label{eq:si_Zfull_from_wedge}
\end{equation}
In the code this operation is implemented by the integer Miller-index map
$\mathbf G_{mn}\mapsto S_\alpha\mathbf G_{mn}$, stored as an index map over the retained $G$-star basis.

For a target wedge momentum $\mathbf k$, the positive exchange matrix is then computed as
\begin{align}
X_{\mathbf G\mathbf G'}(\mathbf k)
&=
\frac{1}{A_W}
\sum_{\ell=1}^{N_{\mathrm{full}}}
w_\ell
\sum_{\mathbf G_1}
Z^{\mathrm{full}}_{\nu\mathbf G_1}(\mathbf k_\ell)
Z^{\mathrm{full}*}_{\nu,\mathbf G_1-\mathbf G+\mathbf G'}(\mathbf k_\ell)
\nonumber\\
&\quad\times
\widetilde v\!\left[
(\mathbf k+\mathbf G)-(\mathbf k_\ell+\mathbf G_1)
\right].
\label{eq:si_exchange_full_bz}
\end{align}
The shifted index appearing in Eq.~\eqref{eq:si_exchange_full_bz} follows from the selection rule in Eq.~\eqref{eq:si_fock_selection}.
The Hartree--Fock matrix used in the diagonalization is therefore
\begin{equation}
h_{\mathbf G\mathbf G'}(\mathbf k)
=
T_{\mathbf G\mathbf G'}(\mathbf k)
+
V^H_{\mathbf G\mathbf G'}
-
X_{\mathbf G\mathbf G'}(\mathbf k).
\label{eq:si_hf_matrix_for_bs}
\end{equation}

Equivalently, one may avoid forming the explicit full-zone list by writing the same exchange sum as a wedge sum over representatives and their distinct symmetry images,
\begin{align}
X_{\mathbf G\mathbf G'}(\mathbf k)
&=
\frac{1}{A_W}
\sum_w
\sum_{j=1}^{s_w}
w_{w j}
\sum_{\mathbf G_1}
Z^{\mathrm{full}}_{\nu\mathbf G_1}(S_{\alpha_j(w)}\mathbf k_w)
Z^{\mathrm{full}*}_{\nu,\mathbf G_1-\mathbf G+\mathbf G'}(S_{\alpha_j(w)}\mathbf k_w)
\nonumber\\
&\quad\times
\widetilde v\!\left[
(\mathbf k+\mathbf G)-
(S_{\alpha_j(w)}\mathbf k_w+\mathbf G_1)
\right].
\label{eq:si_exchange_wedge_equivalent}
\end{align}
This form is algebraically equivalent to Eq.~\eqref{eq:si_exchange_full_bz}; the production results reported here use the explicit full-BZ form, which is less compact but avoids additional bookkeeping inside the exchange summation.


\section{Interpretation of the maximum of the static dielectric function}
\label{sec:si_qmax_dielectric}

The maximum of $\varepsilon_{\mathrm{HF}}(q,0)$ provides additional information
about the momentum-transfer dependence of screening in the Wigner-crystal
state. We denote by $q_{\max}$ the momentum at which
$\varepsilon_{\mathrm{HF}}(q,0)$ reaches its maximum. This is the momentum
transfer for which the additional electronic screening produced by the
resident Wigner crystal is strongest.

The existence of a maximum at finite $q$ follows from the insulating and
crystalline character of the Hartree--Fock solution. In the long-wavelength
limit, the Wigner crystal does not provide a metallic intraband response.
Accordingly, the density response is suppressed and
$\varepsilon_{\mathrm{HF}}(q,0)\rightarrow1$ as $q\rightarrow0$. At large
$q$, the Keldysh interaction is weaker and the density-matrix form factors
entering the Hartree--Fock polarizability become less effective. The strongest
static screening therefore occurs at an intermediate momentum, where the
interaction strength, the interband density matrix elements, and the
Hartree--Fock excitation energies combine to produce the largest screening
correction.

For carrier--carrier scattering, momentum transfers close to $q_{\max}$ are
the ones most strongly renormalized by the electronic response of the Wigner
crystal. In this sense, $q_{\max}$ identifies the most efficient
momentum-transfer channel for static screening in the crystalline state. This
does not make $q_{\max}$ a thermodynamic order parameter; rather, it is a
finite-$q$ response scale associated with the self-consistent crystalline
Hartree--Fock state.

The shift of $q_{\max}$ toward smaller absolute momenta with increasing $r_s$
is consistent with the decrease of the reciprocal-lattice scale as the
Wigner-crystal lattice constant increases. For a triangular Wigner crystal, the
magnitude of the first reciprocal-lattice vector is
\begin{equation}
G_1
=
\frac{4\pi}{\sqrt{3}a_L},
\end{equation}
where $a_L$ is the Wigner-crystal lattice constant. Since $a_L$ increases with
$r_s$, the characteristic reciprocal-lattice scale decreases. The finite-$q$
features in $\varepsilon_{\mathrm{HF}}(q,0)$ are therefore naturally displaced
toward smaller absolute momenta in the dilute regime.

\end{document}